\documentclass[10pt,preprint]{aastex}

\usepackage{rotating}

\slugcomment{accepted by ApJS}

\shorttitle{Chemical Enrichment of the Antennae hot ISM}
\shortauthors{Baldi et al.}

\begin{document}

\title{Chemical enrichment of the complex hot ISM of the Antennae Galaxies: I. Spatial and
spectral analysis of the diffuse X-ray emission}


\author{A. Baldi, J.C. Raymond, G. Fabbiano, A. Zezas, A.H. Rots} 
\affil{Harvard-Smithsonian Center for Astrophysics, 60 Garden St, Cambridge, MA 
02138}
\email{abaldi@cfa.harvard.edu; jraymond@cfa.harvard.edu; pepi@cfa.harvard.edu; 
azezas@cfa.harvard.edu; arots@cfa.harvard.edu}

\author{F. Schweizer}
\affil{Carnegie Observatories, 813 Santa Barbara St, Pasadena, CA 91101}
\email{schweizer@ociw.edu}

\author{A.R. King}
\affil{Theoretical Astrophysics Group, University of Leicester, Leicester LE1
7RH, UK}
\email{ark@astro.le.ac.uk}

\and

\author{T.J. Ponman}
\affil{School of Physics \& Astronomy, University of Birmingham, Birmingham 
B15 2TT, UK}
\email{tjp@star.sr.bham.ac.uk}
%
%

\begin{abstract}
We present an analysis of the properties of the hot interstellar medium (ISM) in
the merging pair of galaxies known as The Antennae (NGC 4038/39), performed using
the deep, coadded $\sim$411 ks {\it Chandra} ACIS-S data set. These deep X-ray
observations and {\it Chandra}'s high
angular resolution allow us to investigate the properties of the hot ISM
with unprecedented spatial and spectral resolution.
Through a spatially resolved spectral analysis, we find a 
variety of temperatures (from 0.2 to 0.7 keV) and $N_H$ (from Galactic to $2\times10^{21}$ 
cm$^{-2}$). 
Metal abundances for Ne, Mg, Si, and Fe vary dramatically throughout the ISM from
sub-solar values ($\sim$0.2) up to several times solar. 
\end{abstract}



\section{Introduction}

The Antennae, dynamically modeled by Toomre \& Toomre (1972) and Barnes
(1988), are the nearest pair of colliding galaxies
involved in a major merger (D = 19~Mpc for H$_o = 75~\rm km~s^{-1}~Mpc^{-1}$).
Hence, this system provides a unique opportunity for getting the most
detailed look possible at the consequences of a galaxy merger as
evidenced by induced star formation and the conditions in the ISM.
Each of the two colliding disks shows rings of giant H~II regions and bright
stellar knots with luminosities up to $M_V \sim -16$ (Rubin, Ford, \& D'Odorico 
1970),
which are resolved with  the  {\it Hubble Space Telescope (HST)}
into typically about a dozen young star clusters
(Whitmore \& Schweizer 1995; Whitmore et al.\ 1999). These knots coincide with 
the peaks of
H$\alpha$, 2.2$\mu$, and 6-cm radio-continuum emission  
(Amram et al.\ 1992; Stanford et al.\ 1990; Neff \& Ulvestad 2000),
indicating an intensity of star formation in each single region exceeding
that observed in 30 Doradus.  CO aperture synthesis maps reveal major
concentrations of molecular gas, including a $\sim$2.7$\times 10^9
M_{\odot}$ concentration in the region where the two disks overlap
(Stanford et al.\ 1990; Wilson et al.\ 2000). This concentration in the overlap
region is coincident with the bright emission peaks seen in {\it SCUBA} 450$\mu$
and 850$\mu$ maps and {\it ISOPHOT} 60$\mu$ and 100$\mu$ observations (Haas et
al.\ 2000), which show
the simultaneous presence of warm dust ($T\approx30$ K), usually observed in
starbursts of luminous IR galaxies, and cold dust ($T<20$ K) more typical
for quiet galaxies and extremely dense clouds.
A K-band study derives ages for the star clusters ranging from 4 
to 13~Myr, and yields high values of extinction ($A_V \sim 0.7 - 4.3$~mag;
Mengel et al.\ 2000, 2005). Many of the youngest star clusters lie in the CO-rich
overlap region, where even higher extinction is present ($A_V \la 100$ mag),
implying that they must lie in front of most of the gas (Wilson et al.\ 2003). Some
of them are found as far as 2~kpc from detectable CO regions, leading to the
conclusion that either they formed from clouds less massive than 
$5\times10^6$ $M_\odot$ or they have already destroyed their parent molecular 
cloud, or perhaps both.

The presence of an abundant hot ISM in The Antennae was originally suggested by
the first {\it Einstein} observations of this system (Fabbiano \&
Trinchieri 1983), and has since been confirmed with several
major X-ray telescopes ({\it ROSAT}: 
Read, Ponman, \& Wolstencroft 1995, Fabbiano, Schweizer, \& Mackie 1997; and 
{\it ASCA}: Sansom et al.\ 1996).
The {\it ROSAT} HRI image was used in a recent multiwavelength study of The
Antennae by Zhang, Fall, \& Whitmore (2001), which suggests feedback between 
star clusters and the interstellar medium (ISM).
The first {\it Chandra} ACIS (Weisskopf, O' Dell \& van Speybroeck
1996) observation of The 
Antennae in 1999 (Fabbiano, Zezas \& Murray 2001;
Fabbiano et al.\ 2003, hereafter F03) gave us for the first time a detailed 
look at this hot ISM, revealing a complex, 
diffuse and soft emission component responsible for about half of the 
detected X-ray photons from the two merging galaxies.
The  spatial resolution of {\it Chandra} is at least
10 times superior to that of any other X-ray observatory, allowing us to 
resolve the emission on physical scales of $\sim$75~pc for D$=$19~Mpc, and to 
detect and subtract individual point-like sources (most likely X-ray binaries; 
see Zezas et al.\ 2002).
At the same time, ACIS
allows us to study for the first time the X-ray spectral properties of these 
emission regions, providing additional important constraints on their nature.

In the first detailed {\it Chandra} study of the hot ISM, F03,
using the 1999 data, presented a ``multi-color''
X-ray image that suggests both extensive absorption by the dust in this
system, peaking in the overlap region, as well as variations in the
temperature of different emission regions of the hot ISM.
These conclusions were confirmed by spectral fits to the data extracted 
from different regions, which suggested that at least two thermal components with
temperatures of $\sim$0.3 and $\sim$0.7~keV are present in the luminous regions
of the diffuse emission. The parameters derived for the hot ISM suggested that
in the nuclear regions of the two colliding galaxies the mass of the hot ISM
is $\sim$10$^5$--10$^6 M_{\odot}$, much smaller than that in
molecular gas (Wilson et al.\ 2000), but that the thermal pressures of the
two phases of the ISM are comparable, suggesting equilibrium.
Comparisons with H$\alpha$, radio-countinuum (Neff \& Ulvestad 2000), and 
HI data (Hibbard et al.\ 2001)
suggested a displacement of the diffuse X-ray emission on the northen side
of the system, consistent with the effect of ram pressure from the cold ISM (H~I)
raining onto the hot gas in the northern star-forming arm.
F03 did not make any conclusive statement on the metal abundance of the hot ISM.
Metz et al. (2004) analyzed the 1999 data together with the second and
third observations, concentrating on regions of
high star-formation rate, and found that super star clusters
play a significant role in heating the ISM, but that the
amount of hot interstellar gas is not directly proportional
to the cluster mass.

While this first {\it Chandra} data set demonstrated the richness of the ISM
in The Antennae, the number of detected photons was such that 
the detailed small-scale morphology and spectral properties could not be 
fully explored.
We now have completed a deep monitoring observing campaign of The Antennae with
{\it Chandra} ACIS, that has produced a very detailed and rich data set. 
Our first look at these data revealed an intricate diffuse emission,
with clear signatures of strong line emission in some regions, pointing to high
metal abundances of the hot ISM  (Fabbiano et al.\ 2004). In the present follow-up paper, we
report a detailed analysis of this diffuse emission. The paper is organized as 
follows.
In \S2 we describe the {\it Chandra} ACIS-S observations of The Antennae
and the procedures followed to screen and prepare the data for analysis. The 
creation of a broad-band mapped-color image of the diffuse emission in The 
Antennae is described in \S3, 
while the preparation of a mapped-color line-strength image---useful for
detecting differences in the spatial distribution of the
metals---is described in \S4. In \S5 we present the spectral analysis 
performed after subdividing the distribution of
diffuse emission into 21 regions, following the indications from the mapped-color 
images. We then discuss
the adopted spectral models and the method used for computing the errors, 
present the results of the spectral fits, and
compare our findings with previous results obtained by Sansom et al.\ (1996) 
and F03. 
We summarize the results of our analysis in \S6.
In a companion paper (Baldi et al., submitted) we discuss the implications of these
results for our understanding of the physical conditions and metal enrichment in the
hot ISM.

\section{Observations and Data Preparation}\label{obs}

NGC 4038/39 were observed with {\it Chandra} ACIS-S seven times during the period 
between 1999 December and 2002 November, for a total of $\sim$411 ks 
(Table~\ref{obslog}). 
The satellite telemetry was processed at the {\it Chandra} X-ray Center (CXC) 
with the Standard Data Processing (SDP) procedure, to correct for the motion 
of the satellite and to apply instrument calibrations. The data used in 
this work were processed in custom mode with the version R4CU5UPD6.5 of 
the SDP, to take advantage of improvements in processing software and 
calibration, in advance of data reprocessing. Verification of the data 
products showed no anomalies.
In addition, we have reduced
positional errors by using stars in the field detected in X-rays (see Zezas et al. 2002).
This revised astrometry is consistent with absolute 
source positions with 1-$\sigma$ errors of $0\farcs6$. We have used 
the latest and best {\it Chandra} positional information for our comparison 
with other positional information.
The data products were then analyzed with 
the CXC CIAO v3.0.1 software and XSPEC package. CIAO Data Model tools were 
used for data 
manipulation, such as screening out bad pixels and producing images 
in given energy bands. The latest release of the ACIS CCD calibration 
files (FEF) was used for the analysis.\\

The data were screened to exclude 
the two ``hot'' columns present in ACIS-S3 at chip-x columns 512 and 513 
(at the b and c node boundaries). No screening to remove high background data 
was necessary: the light curve 
extracted from an area of 7 arcmin$^2$ around the observation aimpoint
(excluding the point sources)
showed a constant count rate of $\sim$0.15 counts s$^{-1}$ for every pointing
but the last one (ObsID: 3041), during which the background rate rose to 
$\sim0.45$ counts s$^{-1}$ for about
25 ks. However, this effect is relatively mild (at worst
three times the ``quiet'' background) and on balance the removal of 1/3 of this
observation would be more detrimental to the data analysis than the slightly increased
uncertainties deriving from a larger background level.\\

\begin{figure}[H]
\epsscale{0.80}
\plotone{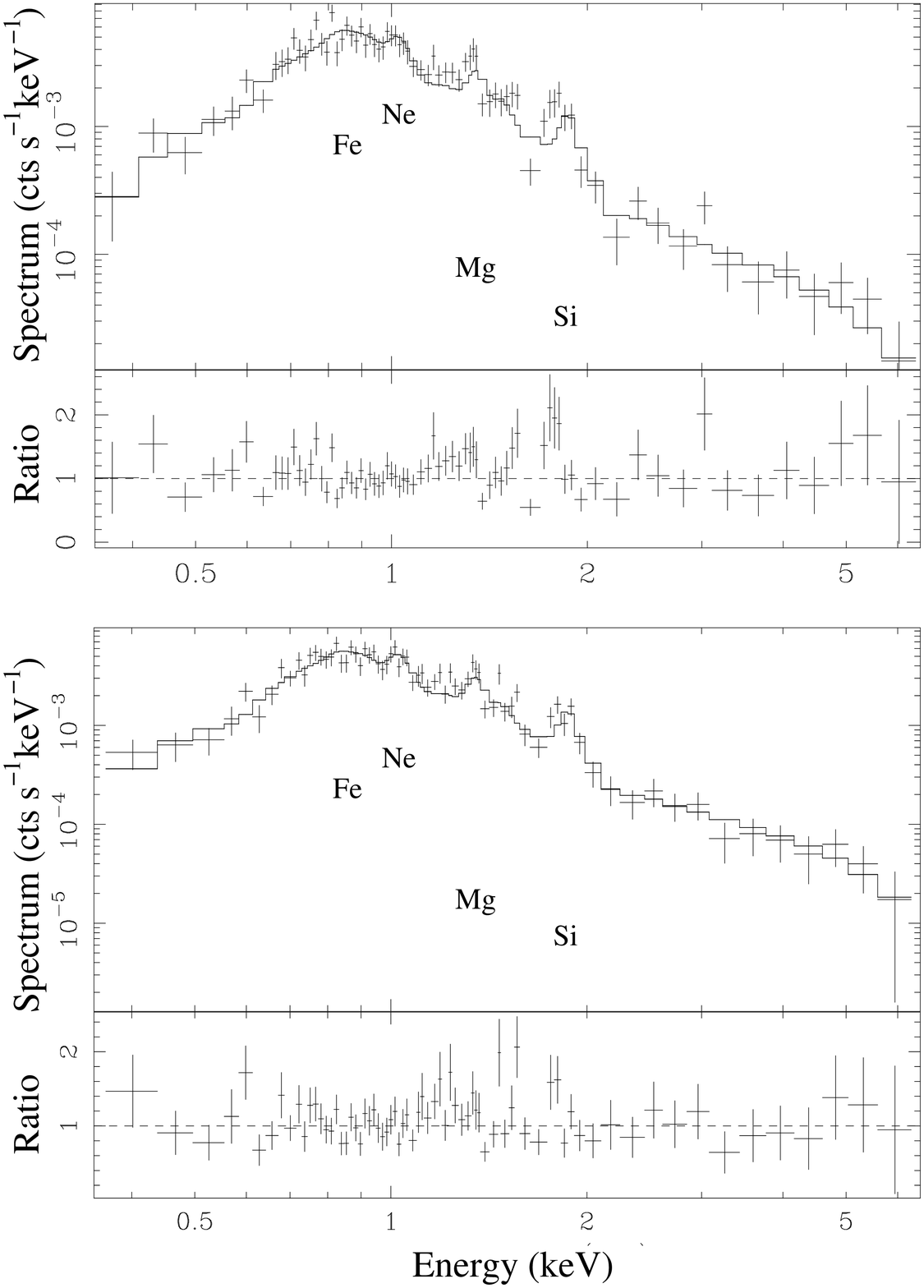}
\caption{{\it Chandra} ACIS-S spectrum of a prominent emission line region of the Antennae hot ISM,
before applying the $tgain$ correction (top) and after applying the $tgain$ correction
(bottom). The spectrum and the ratio with respect to a thermal+power-law 
model (XSPEC: {\em wabs(vapec+powerlaw)}) are plotted.
\label{lineshift}}
\end{figure}

We applied corrections to account for the temporal evolution
of the ACIS spectral response, caused by changes in the 
Charge Transfer Inefficiency (CTI) and in the electronic 
gain of the CCDs. This is a sizeable effect in our case, which we corrected for
by applying the time-dependent gain correction described in
http://cxc.harvard.edu/contrib/alexey/ tgain/tgain.html.
Figure~\ref{lineshift} shows a spectrum
of a diffuse emission region, plotted before and after the gain correction.
A shifting in the emission lines present in the spectrum is evident,
with consequences on the quality of the spectral fits (see
residual plots in Fig.~\ref{lineshift}).\\
The entire data set was then co-added to produce the image 
of the central part of
The Antennae shown in Figure~\ref{totimage}.

\section{The mapped-color image of the diffuse ISM}\label{truecol}

Figure~\ref{totimage} shows that complexly structured diffuse X-ray emission
is clearly present in The Antennae. Previous work on the 1999 observation 
alone has shown that this diffuse emission can be related to a hot ISM with
varying spatial and spectral properties (Fabbiano et al.\ 2001; F03).
However, our much deeper data show a richness of detail in this diffuse 
emission that was not visible in the 1999 data alone.\\

To study this diffuse emission, we followed the procedure of 
F03 to first subtract the point-like X-ray sources
from the data, fill the resulting holes by interpolating the surrounding
diffuse emission, and then produce a mapped-color image of the hot ISM.
First, from the data we extracted images in three different energy bands: 
0.3--0.65 keV, 0.65--1.5 keV and 1.5--6 keV. 
The middle energy band was chosen to encompass the emission from the Fe-L
blend. The 0.3 keV low-energy boundary is the lowest energy within which 
ACIS is well calibrated.
The upper energy boundary was set at 6 keV because at energies $>$6 keV
the diffuse emission becomes background-dominated.
Then, we used the CIAO tool $wavdetect$ to generate a list of sources 
detected
at a threshold of $3\sigma$ in the co-added exposure, using data in 
the total 0.3--6~keV energy band
within the central 2\arcmin\ of the image. From this source list, we removed 
the extended
sources listed in Zezas et al. (2002) plus a 
few additional slightly-extended sources identified by visual inspection of the data.
This culling left a total of 127 point-like sources,
the positions of which are shown in Figure~\ref{totimage}.\\ 

We then used the CIAO tool {\em dmfilth} to subtract these 127 sources from the data.
This tool cuts out the point sources and fills in 
the holes corresponding to the source positions with values interpolated from 
the surrounding background regions.
We adopted as source
regions the $3\sigma$ best fit ellipses computed by $wavdetect$. As
background regions we adopted elliptical annuli having the source region 
ellipse as inner boundary
and an ellipse with the same orientation, but with double size
semi-minor and semi-major axes, as outer boundary.  Areas in the background 
regions where 
unrelated sources happened to fall were excluded from the calculation of 
the background. 
We used an
interpolation technique in which pixel values in the source regions are
sampled from the distribution of pixel values in the background regions
({\em dist} option in {\em dmfilth}), since
the Poisson interpolation in the current version of 
{\em dmfilth} does not work for
low background regions ($<0.10$ cts/pixel) such as those in the outer parts of 
The Antennae.\\

The three diffuse-emission images at different energies were then adaptively smoothed,
using  {\em csmooth}. This process preserves the high-significance small-scale details 
of the image, while enhancing more extended features of low surface brightness. 
The same 
smoothing scales were used for the three images, based on those computed during 
the smoothing of the 0.65--1.5 keV image, which is the one with the highest 
signal-to-noise ratio. 
For the smoothing, we used a 
Gaussian kernel of varying size (from 1 to 40 pixels). The maximum significance 
was set to 5$\sigma$, while the maximum smoothing scale corresponds to a 
minimum 
significance 
of $\sim$2.4$\sigma$ for the 0.6--1.5 keV image.
The resulting smoothed images were then
corrected for exposure (i.e., flat-fielded) by dividing them by the appropriate 
exposure maps, smoothed using the same scale
as the image.
The exposure maps were generated for
a thermal emission model (Raymond \& Smith 1977) with a temperature of 0.8 keV.
The resulting exposure-corrected and adaptively-smoothed images, shown in
Figure~\ref{3color}, were then combined 
(using the tool $dmimg2jpg$)
to create the ``mapped-color'' image of the diffuse emission, shown in 
Figure~\ref{truecolor}.\\

In Figure~\ref{truecolor}, red denotes 
softer X-ray emission regions (0.3--0.65 keV), while blue denotes 
harder emission regions (1.5--6 keV). Green identifies emission in the middle 
band (0.65--1.5 keV).
In the following discussion, we will refer to the counts in these bands as $S$ 
(for soft), $M$ (medium), and $H$ (hard).
The color bars in this figure quantify the distribution of the pixel 
values (from minimum to maximum) in each band.\\

By comparison with the mapped-color image obtained from the first dataset (ObsID: 
0315; F03),
the new, deeper image shows an impressive level of detail.
Spatial and spectral structures are visible on different spatial scales,
from the smaller-scale clumps in the inner parts of the optical merging 
galaxies to the larger-scale regions of low surface brightness .\\

Since we have three color-coded energy bands, the resulting pixel color will be 
ideally distributed in a
tridimensional ``color cube'' whose axis values are proportional to the number of 
counts in
a given band (see right panel of Figure~\ref{colorbars}). 
We have ``sliced'' this cube to obtain a two-dimensional representation
that relates the pixel color to the hardness ratios $HR1$ and $HR2$, defined 
as:
$$
HR1=\frac{M-S}{M+S};
\:\:\:
HR2=\frac{H-M}{H+M}.
$$
Each of these slices represents a plane with the same value of hard-band 
counts (see Figure~\ref{colorbars}),
with soft-band counts increasing along the horizontal axis and medium-band counts 
increasing vertically.
For this chosen value of hard counts, each color in the slice can be uniquely 
related to a value of $HR1$.
These values are written in the bottom slice. The same values will apply to 
each slice.
$HR2$ does not depend on the amount of soft counts, so we display a column of 
$HR2$ values for each slice, along the ``medium-band'' axis.\\

\begin{figure}[H]
\epsscale{0.90}
\plotone{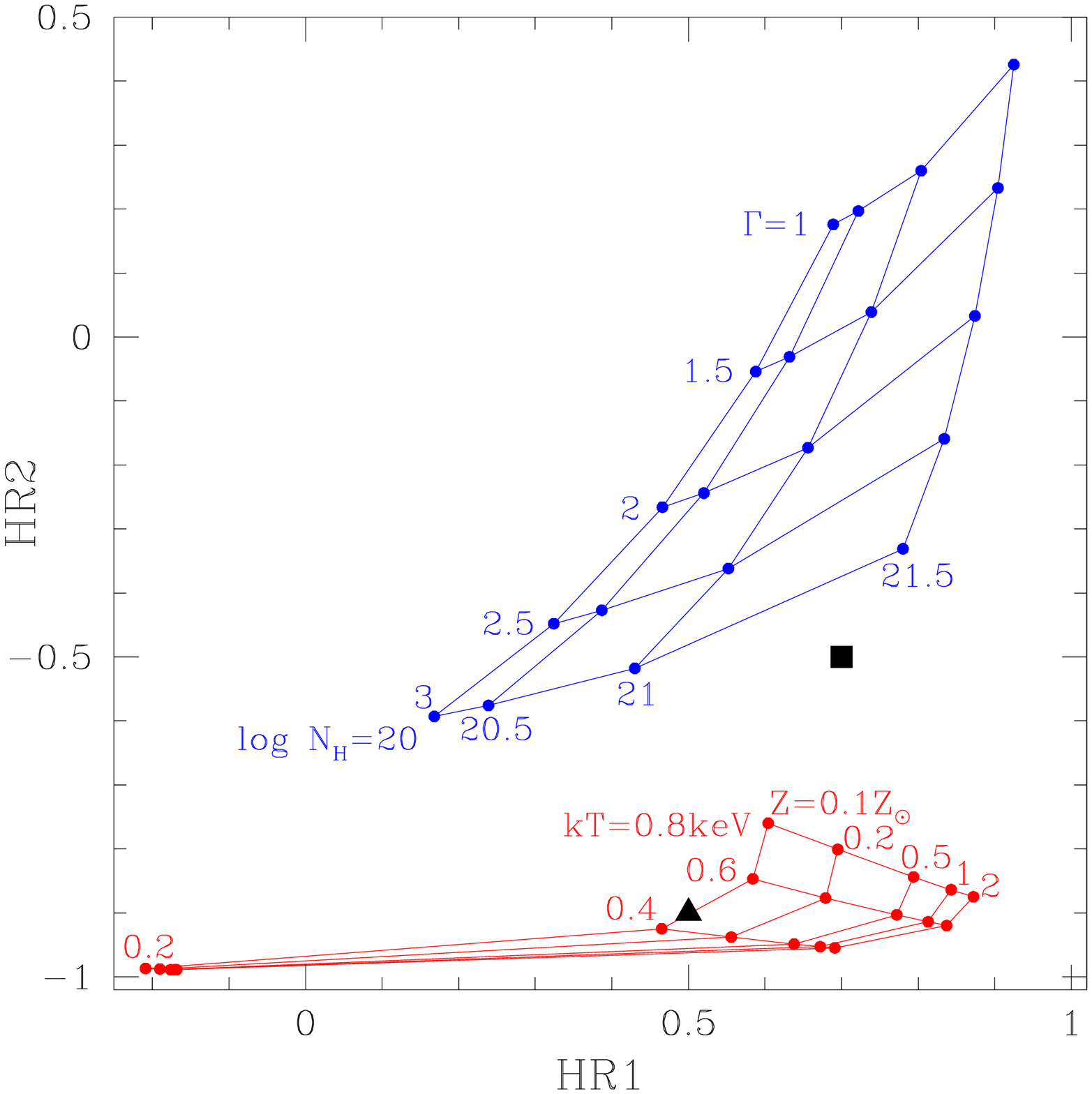}
\caption{X-ray color-color diagram of $HR1$ vs.\ $HR2$. The blue grid indicates
the regions occupied in the case of a power-law model, varying the intrinsic
absorption $N_H$ and the photon index $\Gamma$. The red grid represents the
loci for a thermal model, varying the temperature $kT$ and the metallicity
$Z$. The filled square and triangle indicate the positions in the
diagram of the regions described in $\S$3.
\label{hrgrid}}
\end{figure}

It is possible to also give a physical meaning to these numbers, 
computing the values of $HR1$ and $HR2$ in two cases: an absorbed Raymond \& 
Smith (1977) thermal model and an absorbed power-law
model. For the thermal model we vary the temperature from 0.2 keV to 0.8 keV 
and the metallicity from 0.1 solar 
to 2 times solar. In the power-law case the photon index $\Gamma$ ranges from 1 
to 3 and the intrinsic absorption from $10^{20}$ to
$10^{21.5}$ cm$^{-2}$. In both cases we considered also a Galactic $N_H$ 
component fixed at the 
line-of-sight value toward The Antennae ($\sim3.4\times10^{20}$
cm$^{-2}$; Stark et al.\ 1992). The values of $HR1$ and $HR2$ are plotted
in the X-ray color-color diagram shown in Figure~\ref{hrgrid}, and they can be
directly related to the labels in Figure~\ref{colorbars}. The red grid refers 
to the thermal model, while the blue grid represents
the power-law model.
It is likely that a simple single-temperature thermal model or an absorbed 
power-law model may not be a good fit to
the data.  However, the figure gives an indication of the characteristics
of the X-ray emission from different regions.\\ 

For example, the blue region on the bottom-left of Fig.~\ref{truecolor}
corresponds roughly to $HR1\sim0.7$ and $HR2\sim-0.5$.
Looking at Figure~\ref{hrgrid}, these hardness ratios---marked by a black 
square---are located at a point in the diagram between the
absorbed-power-law model and the purely thermal model, possibly
indicating the presence of unresolved binary
emission in this region.
On the other hand the orange region on the right of Fig.~\ref{truecolor}
corresponds roughly to
$HR1\sim0.5$ and $HR2\sim-0.9$ 
(black triangle in Figure~\ref{hrgrid}). These values are typical
of purely thermal emission.\\

\section{The Line-Strength Map}

As Figure~\ref{lineshift} illustrates, the spectra extracted from different 
regions of the diffuse emission
frequently show emission lines. In particular, O+Fe+Ne emission
(0.6--1.16~keV) is often present, 
together with Mg-XI (1.27--1.38~keV) and Si-XIII (1.75--1.95~keV) lines. Using 
these features, we generated another color map (see Fabbiano et al.\ 2004)
that gives a qualitative representation of the presence of emission lines from
Iron, Silicon, and Magnesium in different regions of the hot ISM.
This map has proven very useful for identifying regions of strong apparent
line emission and for selecting spectrally similar regions for a proper
spectral analysis.

The line-strength map, shown in Fig.~\ref{regions21}a, was constructed as follows.
First, we created images in three bands
encompassing emission from O+Fe+Ne (0.6--1.16~keV), the Mg-XI
line (1.27--1.38~keV), and the Si-XIII line (1.75--1.95~keV). We also
created an image  including emission in the 1.4--1.65~keV and
2.05--3.05~keV bands, which was used for continuum subtraction. The
line and the continuum bands were 
determined from spectra of the diffuse emission in order to have 
an optimum line/continuum ratio  and minimal
contamination by emission lines, respectively.
In each band we excluded all point-like sources and interpolated
over the holes as described in \S~\ref{truecol}. The O+Fe+Ne
band image,
which contains a much larger number of counts than the other line-band images,
was used to determine the smoothing scales, and was 
adaptively smoothed  to a significance between 3$\sigma$ and
5$\sigma$.  The same smoothing scales were then also applied to 
the other line and continuum images. 
In order to subtract the appropriate continuum from each line image,
the continuum image was rescaled to represent the band continuum in the given 
line. To calculate the weighting factors, 
we extracted the spectrum of the entire diffuse emission of The Antennae and
fitted the continuum (after excluding the energy ranges with line
emission) with a two-component Bremsstrahlung model. Based on this
model we estimated the relative intensity of the continuum in our
continuum band and the three line bands. Each line image was
continuum subtracted after weighting the continuum image by these
weighting factors, and the resulting images were combined to create
a mapped-color line-strength image (Figure~\ref{regions21}a).
In this image, red represents line emission from O+Fe+Ne,
green from Magnesium, and blue from Silicon,
respectively.

Figure~\ref{regions21}a shows that the hot ISM of The Antennae has a complex 
emission-line structure.
Silicon emission (blue) appears pervasive in most of the hot ISM, with
Magnesium lines (green) also present in various relative amounts and O+Fe+Ne
emission (red) prevalent in some regions.
Other regions (black) show only very weak or no line emission.

\section{Spectral Analysis}\label{specanal}

The mapped-color X-ray image (Figure~\ref{truecolor}) and the line-strength map 
(Figure~\ref{regions21}a)
demonstrate that the diffuse emission of the hot ISM is rich in spatial and
spectral features, and that
the line emission in the X-ray spectra is pervasive and varied. These figures 
suggest a temperature 
structure in the hot ISM, possibly the effect of varying absorption on the 
emission (see also F03), and of its metal enrichment.

\subsection{Region Selection and Extraction of the X-ray Spectra}\label{selreg}

To put these results on a more 
rigorous and quantitative footing, we have performed a spectral analysis of 
different regions
of the hot ISM, using
Figures~\ref{truecolor} and \ref{regions21}a as guides for the selection 
of these regions.
We first identified 17 separate regions for analysis, based on the morphology 
of the diffuse
X-ray emission (Fig.~\ref{regions21}a). However, the line-strength map suggested 
marked abundance differences within some of these regions. In particular, 
Figure~\ref{regions21}a 
shows that regions 4, 6, 
8, and 12 have a complex metallicity structure. For example, in Region 4 
we can identify a metal-rich and a metal-poor area; in regions 6, 8, and 12
we see clearly defined sub-areas with yellow 
(O+Fe+Ne and Mg) and  red (O+Fe+Ne) emission. These four regions also have a 
very high number
of detected counts, making it possible for us to subdivide them further on the 
basis of their line
features. We subdivided each of these four regions into two parts, as shown in  
Figure~\ref{regions21}b. 
Although Figure~\ref{regions21}a suggests spatially complex metal distributions also
in other regions (5, 13, 14 and 15), these regions have too
few counts to be subdivided further. The described process identified a total of
21 regions for separate spectral analysis.

From these 21 regions, we excluded all the 
point source $3\sigma$ areas described
in \S~\ref{truecol} and shown in Figure~\ref{totimage}. We then extracted
background counts representative of the field from three circular source-free
regions well outside the galaxies, but within 
the same CCD (ACIS-S3) in all seven 
observations. These background regions covered a total area of $\sim$4 arcmin$^2$.

All of the data sets under analysis, with the exclusion of 
the first and fifth (ObsID: 315 and 3044, respectively),
were obtained with the same ACIS-S configuration and could then be
treated as a single observation. We therefore merged these data into a single event file
by using the CIAO script $merge$\_$all$. 
ObsID 315 and 3044, however, needed individual handling. 
ObsID 315 was obtained in
1999 December, when the low-energy response of ACIS-S was significantly
higher ($A_{eff}\sim500$ cm$^2$, $QE\sim65\%$ at 1 keV)
than at the time of the other six observations (2001 December to 
2002 November).
ObsID 3044 was obtained with a different ACIS-S configuration (different SIM-Z).

We extracted spectra separately from the merged 5-ObsID file and from 
315 and 3044, using the diffuse emission and background areas
described above.  In all cases, 
the CIAO script $acisspec$ was used to extract the region spectra. 
This script invokes the tool $dmextract$ to extract
the spectrum and then uses the tools $mkwrmf$ and $mkwarf$ to create area-weighted
Response Matrix Files (RMF) and Ancillary Response Files (ARF) 
for each region, since they span more than one FITS
Embedded Function (FEF) region.
The spectra derived from the three different event files, 
both from source and background regions, were then summed using the FTOOL $mathpha$.
The response files were
combined with their appropriate weights, using the FTOOLS $addrmf$ and $addarf$.
The resulting spectra were then grouped so as to have at least 20 counts per
energy bin before background subtraction.

For each of the 21 regions used in the spectral analysis the net counts 
(background subtracted) in three different energy bands and in the total 0.3-6.0 keV 
band are listed in Table~\ref{counts}.

\subsection{Spectral Fit Method}\label{specfit}

We analyzed the extracted spectra with XSPEC v11.2.0 (Arnaud 1996),
using combinations of optically-thin thermal emission and power-law
components to fit the data. We used the Astrophysical Plasma Emission Code 
(APEC) thermal-emission model (Smith et al.\ 2001) to represent the thermal 
emission. 
This model is based on the original Raymond \& Smith (1977) code, but 
has been totally revamped to exploit modern computing capabilities 
and the wealth of accurate atomic data.

The real nature of a soft X-ray emitting gas, recently heated as in star-forming
galaxies, is clearly that of a multi-temperature plasma with very complex 
differential emission-measure distributions (see, e.g., Strickland \& Stevens 2000). 
Since the gas is continuously being heated by SN and star winds, it may also not be in 
ionization equilibrium. However, modeling our low S/N spectra with complex
models having more than two temperatures or not assuming ionization equilibrium
would require too many free parameters to 
obtain any meaningful information from them.
For example, although fitting higher S/N spectra of diffuse X-ray emission in nearby
starburst galaxies with two-temperature models (using combinations of several equilibrium
and non-equilibrium models) frequently leads to better fits than fitting with a
single-temperature model (e.g., NGC 253: Strickland et al.\ 2002), 
they almost always lead to unrealistically low abundances, as previously noted by 
Metz et al.(2004).
Therefore, we considered also single-temperature models to fit our data, together
with two-temperature models, to allow us to compare the results obtained from both.
The four models that we used, listed in order of increasing complexity, are:
\begin{itemize}
\item absorbed one-temperature APEC model (XSPEC model: {\em wabs(wabs(vapec))});
\item absorbed one-temperature APEC model plus a power-law component (XSPEC model: {\em wabs(wabs(vapec+pow))});
\item absorbed two-temperature APEC model (XSPEC model: {\em wabs(wabs(vapec+vapec))}); and
\item absorbed two-temperature APEC model plus a power-law component (XSPEC model: {\em wabs(wabs(vapec+vapec+pow))}).
\end{itemize}
Two different absorption components were used in all cases: one fixed at the 
Galactic line-of-sight value toward The Antennae 
($N_H=3.4\times10^{20}$ cm$^{-2}$; Stark et al.\ 1992) and the other one left 
free to vary in order to model the intrinsic absorption within the two galaxies. 
In the two models where a power-law component is included we tried also to fit the 
$N_H$ absorption independently for the power-law and thermal components.
Although the $N_H$ of the power-law component resulted in a value different from the
$N_H$ of the thermal component(s) (on average $\sim10^{21}$ cm$^{-2}$ instead of
$\sim10^{20}$ cm$^{-2}$),
the best fit parameters of the thermal component (mainly $kT$ and $Z$) did not vary
significantly and were fully consistent within the errors.
The abundances of Neon, Magnesium, Silicon, and Iron were all left free to vary. The 
other elements (e.g., Carbon, Nitrogen, and Sulfur) are poorly constrained 
because of gain uncertainties and 
sensitivity to $N_H$. Thus we adopted solar abundances for these elements. 
We tried to thaw the Oxygen abundance, obtaining no clear improvements in the fits and no
significant variations in the best-fit parameters of $kT$ and of $Z$ in the other elements.
Moreover, the poorly costrained Oxygen abundance values were always consistent with solar
values.  Therefore, we decided to adopt a solar abundance also for this element.

We introduced a power-law component into our models to allow for the possible residual
presence of unresolved X-ray binaries too faint to be detected individually. 
Indeed, from an analysis of the entire emission throughout The Antennae we found that
the point sources contribute up to $\sim$85\% of the emission in the 2--5 keV 
range and are essentially the only contributors above 5 keV.
The photon index value was at first frozen at $\Gamma=1.88$, a value 
obtained by fitting the 2--10 keV spectrum of the entire emission from The Antennae, 
including point sources. Afterwards $\Gamma$ was left free to vary to fit the data
in order to test for the presence of substantial discrepancies between the average power law 
determined from the galaxies' global emission and possible power-law components in
individual regions.
We compared the results of the two fits (frozen and thawed $\Gamma$) for the same region
using an F-test, and considered the thawed $\Gamma$ model to be a better
description of the data if $p$, the probability that the improvement in the 
$\chi^2$ statistics is due by chance, was less than 5\%.\\
%

The resulting best-fit temperature(s) $kT$, absorbing column density $N_H$, and
power-law  photon index $\Gamma$ (if a power-law component is necessary) for the
21 regions studied are summarized in Table~\ref{mytable1old}, 
together with the abundances of Neon, Magnesium, Silicon, and Iron relative to
the solar values.
The last column of Table~\ref{mytable1} gives the ``best-fit model''. 
The errors are quoted at $1\sigma$ for one interesting parameters.
The values listed in the table are obtained
from the ``best-fit model'', which is defined as the simpler
model (i.e. with a smaller number of parameters to be fitted) with $\chi^2_{red}<1.4$.
This value of $\chi^2_{red}$
represents a fair boundary between an acceptable fit and a bad fit
for low S/N spectra (0.3--6 keV counts in the range between
$\sim750$ and $\sim6500$). This means that we do not adopt a two-temperature model
if an acceptable fit is already obtained with only one temperature. 
We understand that a single-temperature
and even a two-temperature model are approximations to the 
real physical nature of the X-ray emitting gas, which is likely to be more complex.
However, given the limitations of our data we followed the customary data analysis
practice of choosing the simpler model of two if both models give acceptable representations
of the data. 
Indeed, we found that
only one region does not yield an acceptable fit using a single-temperature spectral model
(Region 4b). Possible systematic biases depending on the model choice are taken into account
as explained in \S~\ref{errorcomp}\ below.
The only region where none of our models yielded a
$\chi^2_{red}<1.4$ (although with only 751 counts in the 0.3--6 keV band) is Region 13, 
for which we chose the model with the lowest $\chi^2_{red}$.
The spectra of the 21 regions, together with the best-fit models and
the data-to-best-fit-model ratios,
are shown in Figures~\ref{spectra1}, \ref{spectra2}, \ref{spectra3}, and 
\ref{spectra4}.\\

\subsubsection{Error computation}\label{errorcomp}

Although quoting the errors at $1\sigma$ for one interesting parameter 
(as the errors quoted in Table~\ref{mytable1old}) is
a common practice in X-ray spectral analysis, it may lead to an 
underestimation of the true uncertainties not only because of possible systematics
but also in the case that some of the fitting parameters are correlated. As previously 
noted by Martin, Kobulnicky \&
Heckman (2002) a degeneracy between metallicity and temperature may exist in the
spectral models, and it is very difficult to solve such ambiguity from low 
resolution CCD spectra like the ones we are dealing with. Indeed we found such
a degeneracy at least in some of our regions. Although it will not be possible
to solve completely this problem from the spectral analysis of ACIS spectra alone, 
it is still possible to adopt a very conservative approach to take into account
both systematics and correlations between fitting parameters.
This approach mainly consists in calculating the errors for the best-fit parameters 
at 1$\sigma$ for all
interesting parameters. The errors on $kT$, $N_H$, and $\Gamma$ were calculated 
from the $\chi^2$ statistics considering all free parameters in the 
``best-fit model'' as interesting. Since the spectral regions occupied by Ne, 
Mg, and Si lines are quite distinct, the errors on the 
abundances for these elements were computed using a different method to reduce 
the number of interesting parameters.
We first attempted to link together all the $\alpha$ elements using the 
type II SNe ratio from the compilation of average stellar yields from
Nagataki \& Sato (1998), in order to reduce the number of free parameters by two. 
However, we were unable to obtain acceptable fits for any of the regions, suggesting
that the elemental ratios are not entirely consistent
with type II SN yield models.
Thus we tried an alternative approach for each individual $\alpha$-element line, 
ignoring from the fit the part of 
the spectrum relative to the other two lines (e.g. for Mg we ignored the part of the 
spectrum occupied by the lines of Ne and Si) and freezing their abundances to 
their best-fit values.
Therefore, in the error computation of the $\alpha$-elements we have two free 
parameters less than in the computation of $kT$, $N_H$, and $\Gamma$.
For deriving the Fe abundance we ignored the part of the spectrum relative 
to all three $\alpha$ elements (freezing them to their best-fit values), 
obtaining three less free parameters than the original model.
The number of interesting parameters was set equal to the number of free 
parameters also for the abundance error estimation. Note that
this is a very conservative approach for the determination of statistical errors, since
the fitting parameters are not completely independent from each other, so that
the number of truly interesting parameters is likely to be lower than the one
adopted. However, statistical errors are not the only uncertainties
present in our measurements, and systematic biases may arise also 
from the spectral model choice, as noted before in \S~\ref{specfit}.
Although our method cannot estimate exactly any possible systematics, it is 
conservative enough that we are confident about the parameter confidence 
ranges so derived.
The best-fit parameters and the errors computed using this method are shown
in Table~\ref{mytable1}. Comparing this table with Table~\ref{mytable1old},
it is clear that the abundances measured using the method described above
are consistent with the measures performed fitting the whole spectrum. The
only noticeable difference comes from the 1$\sigma$ confidence ranges which,
of course, are larger in Table~\ref{mytable1}. In the rest of the paper we
will use the confidence ranges of Table~\ref{mytable1}.

\begin{figure}[h]
\epsscale{1.10}
\plottwo{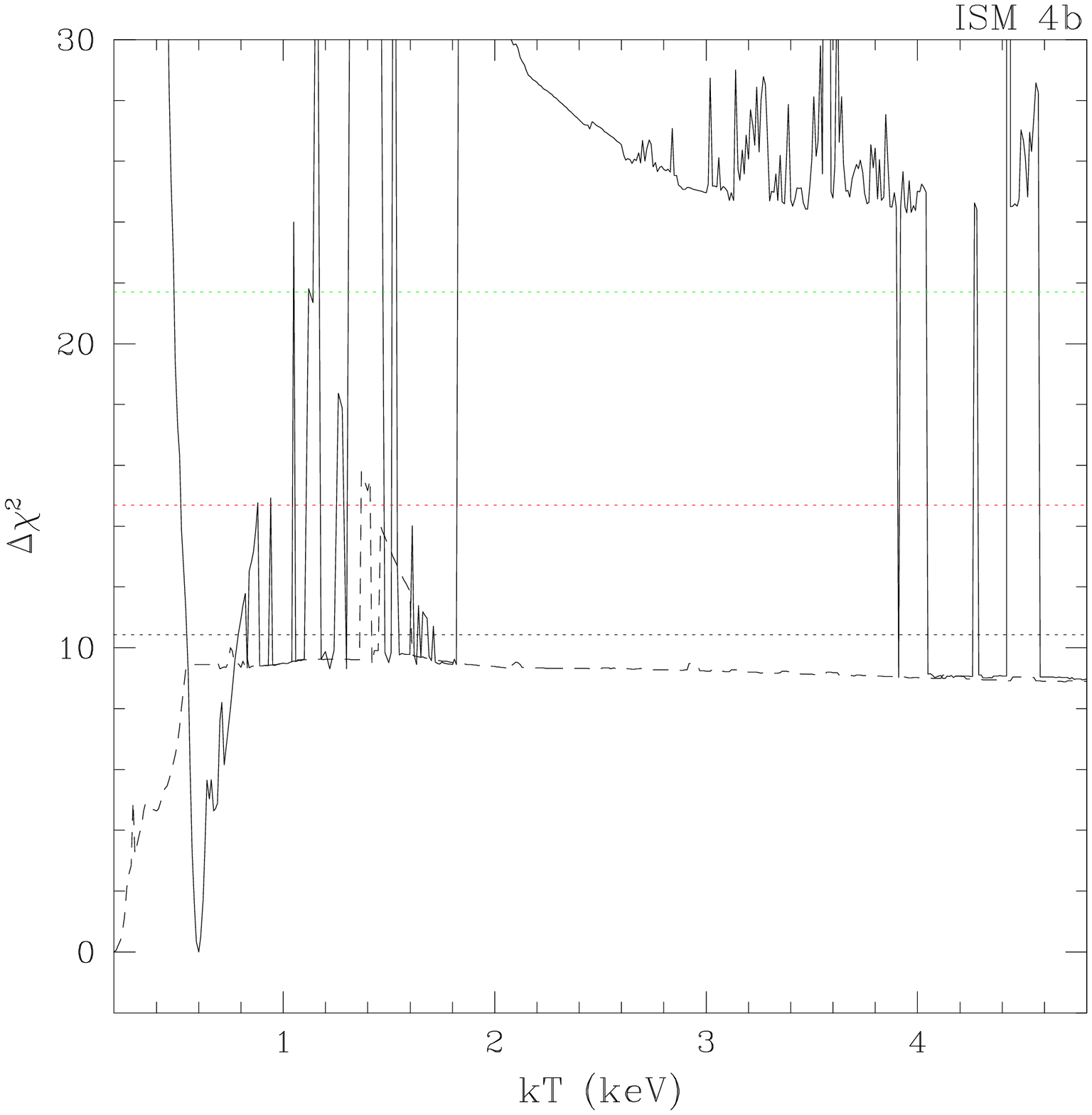}{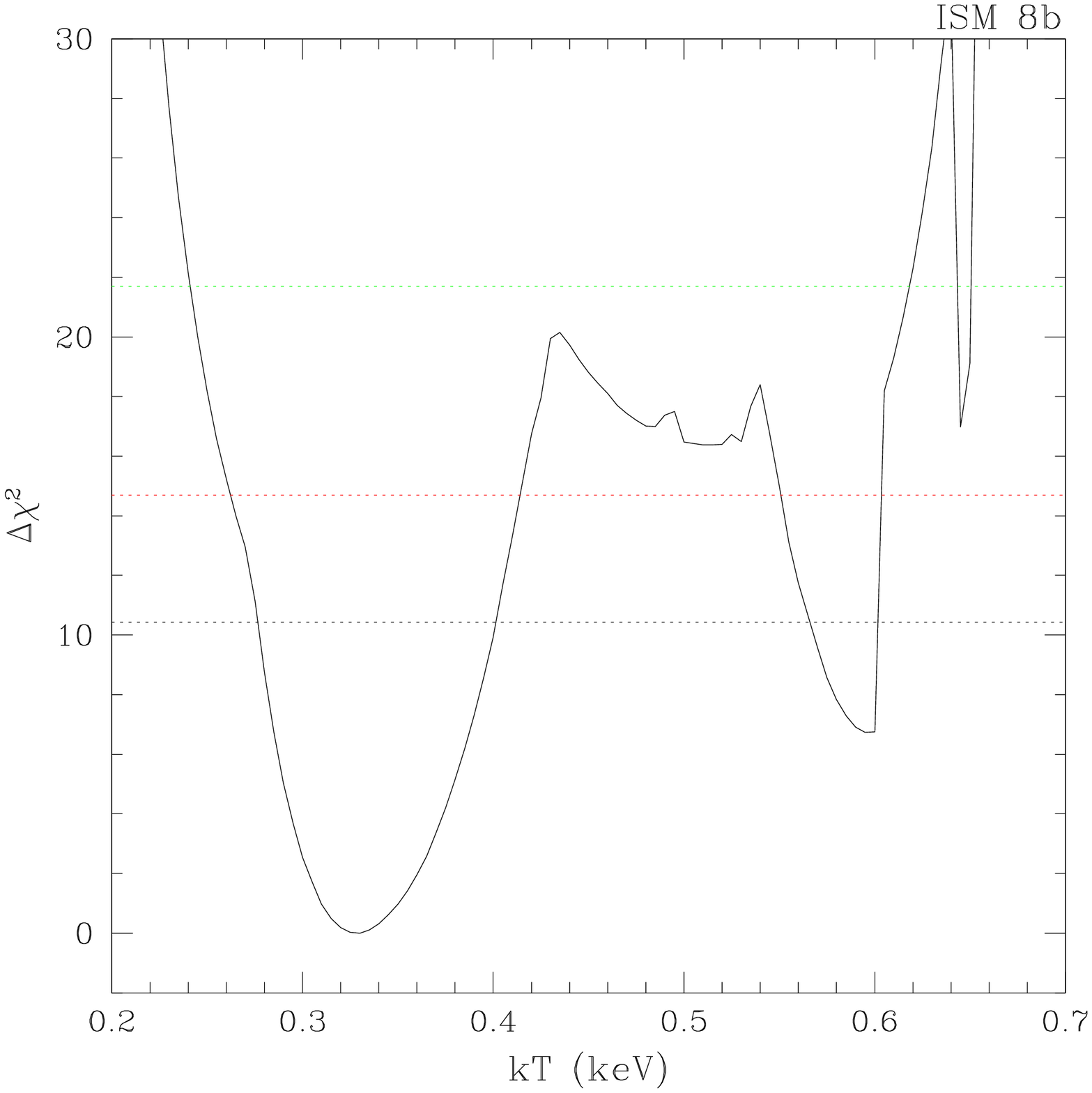}
\plottwo{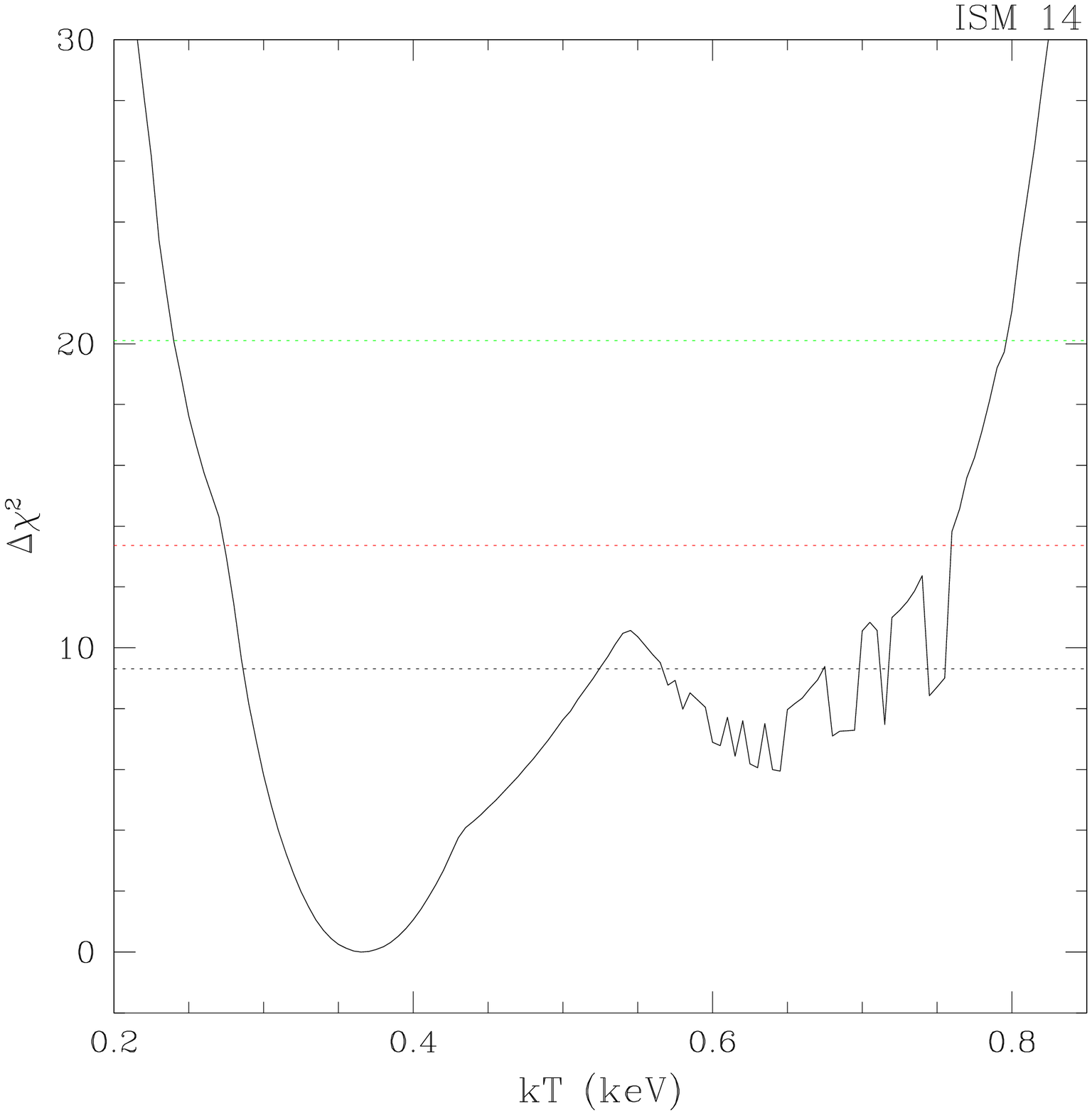}{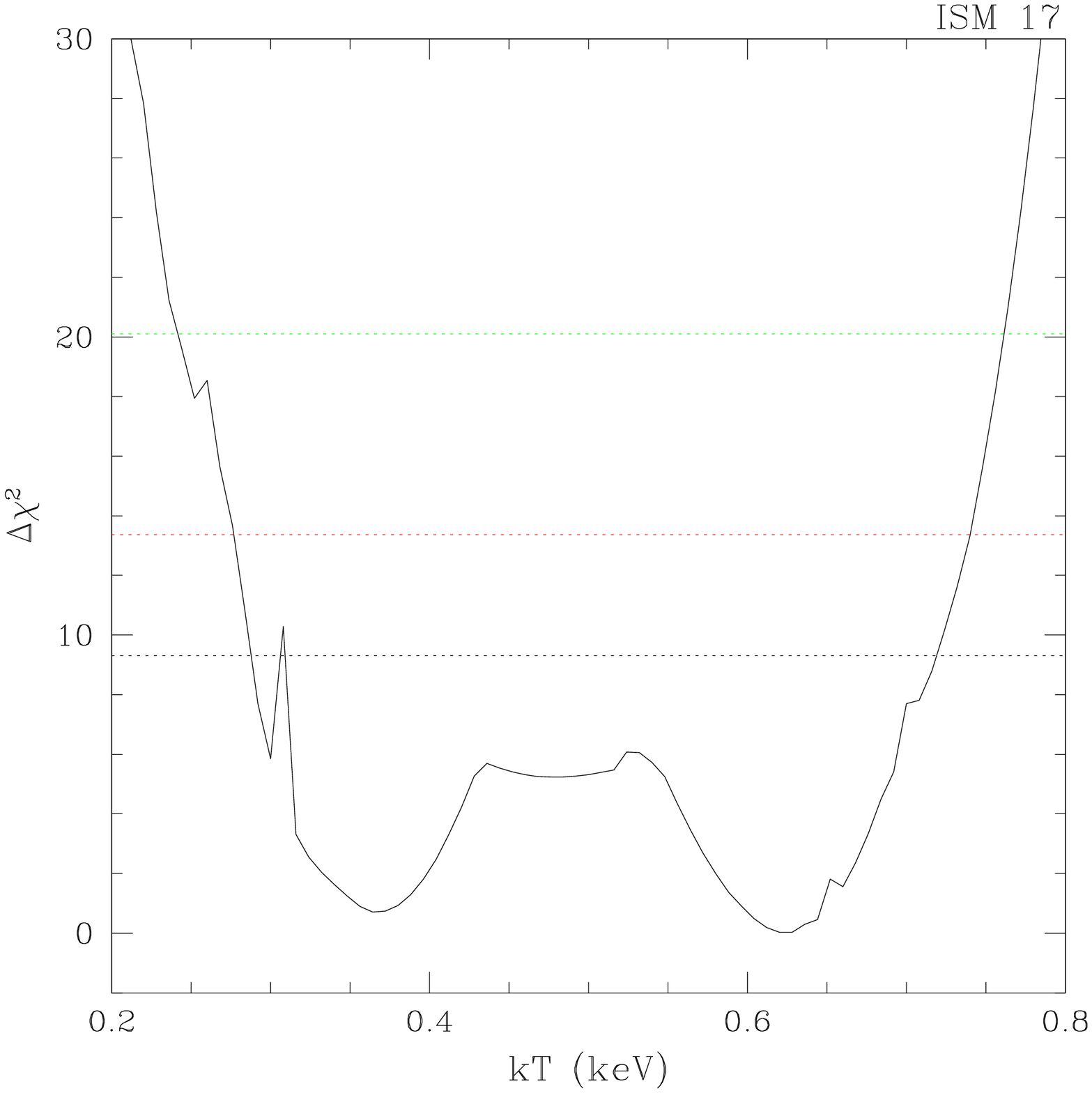}
\caption{Distribution of the $\Delta\chi^2$ values as a function
of the temperature in four of our regions chosen for spectral analysis.
The black dashed line in the upper-left panel represents the $\Delta\chi^2$
distribution as a function of the lower temperature in a two-temperature
best-fit model.
The black dotted line, the red dotted line, and the green dotted line
represent the 68\%, 90\% and 99\% confidence levels, respectively.
\label{chi2kt}}
\end{figure}

\begin{figure}[h]
\epsscale{1.10}
\plottwo{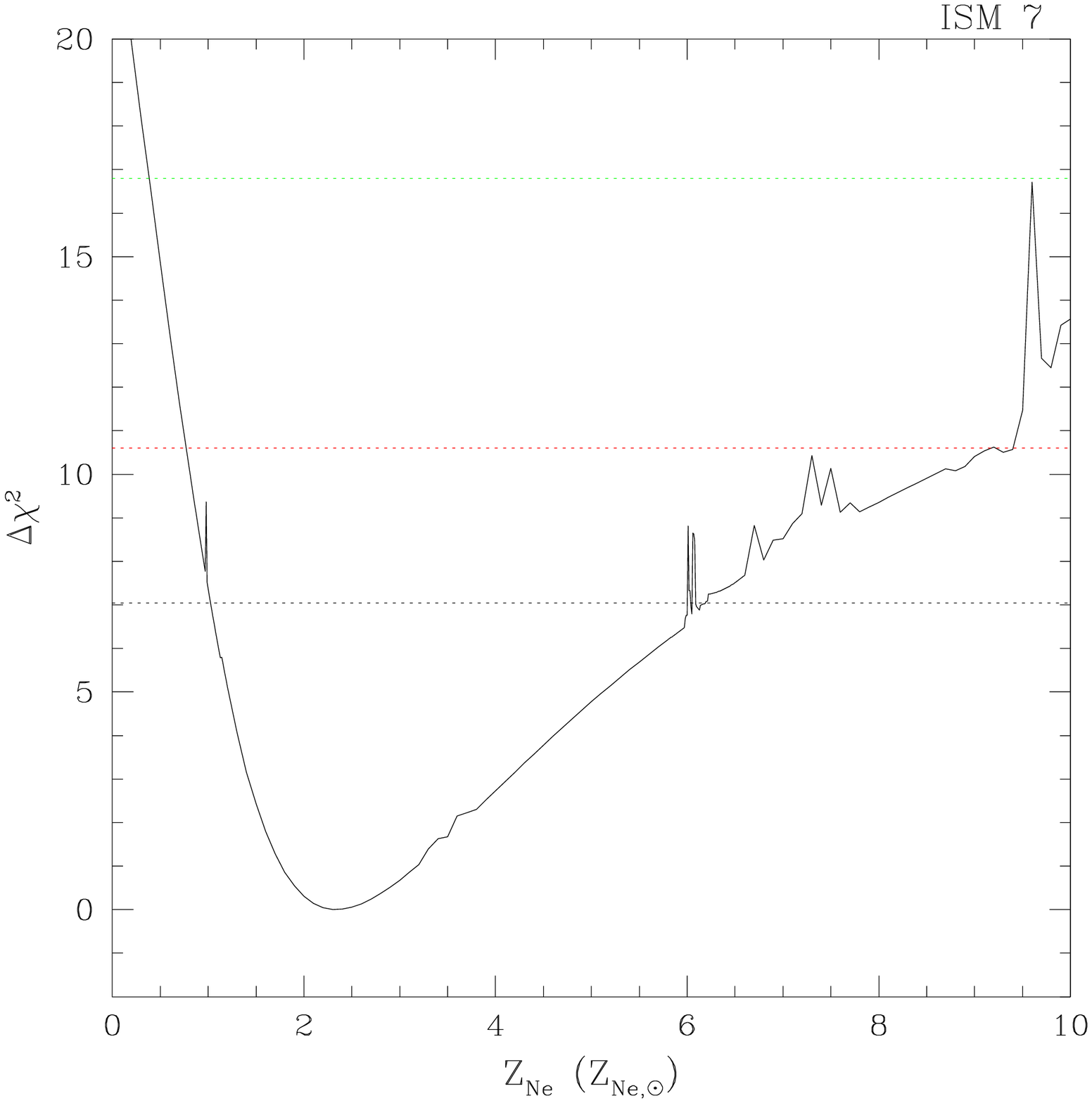}{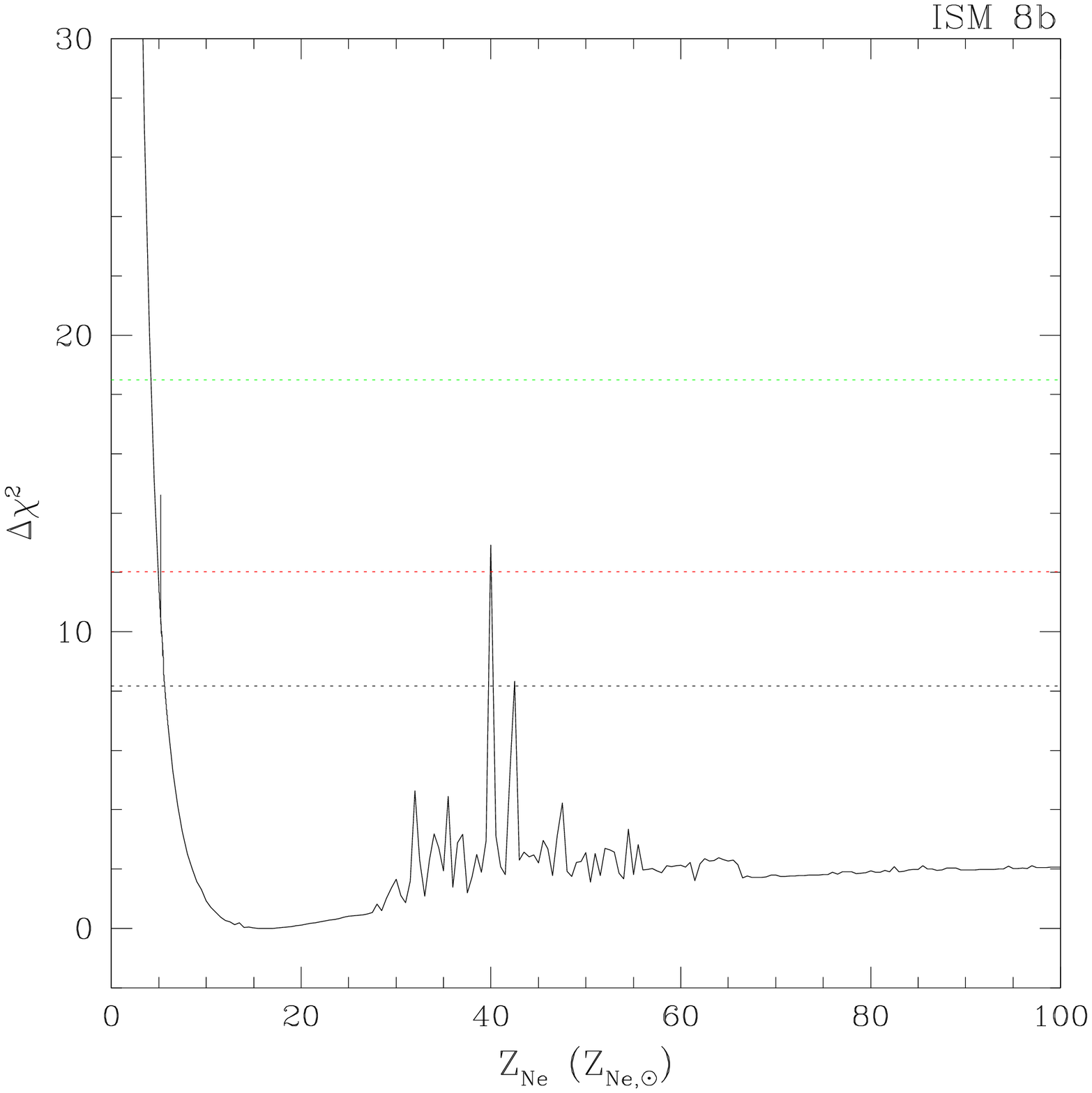}
\plottwo{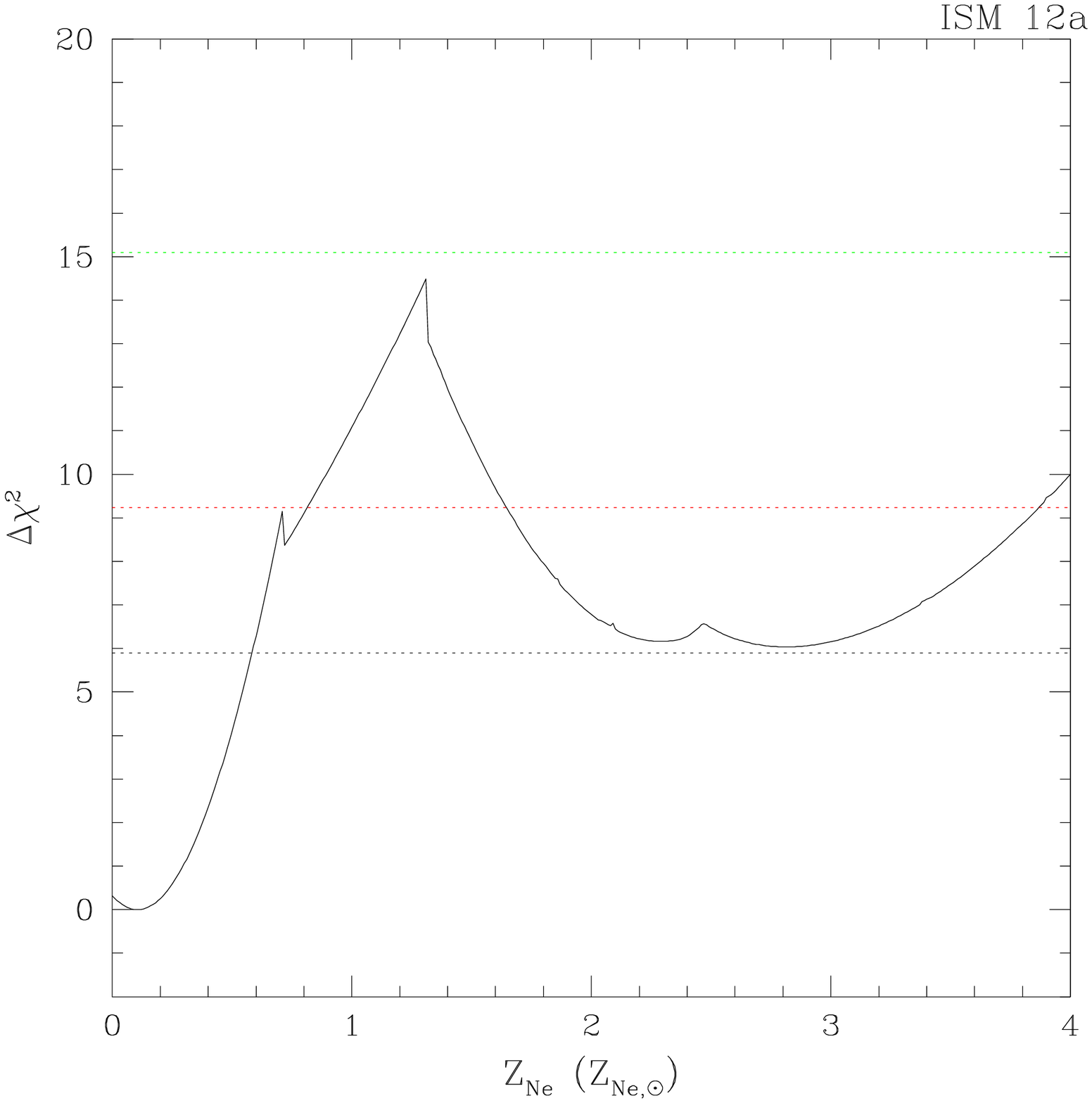}{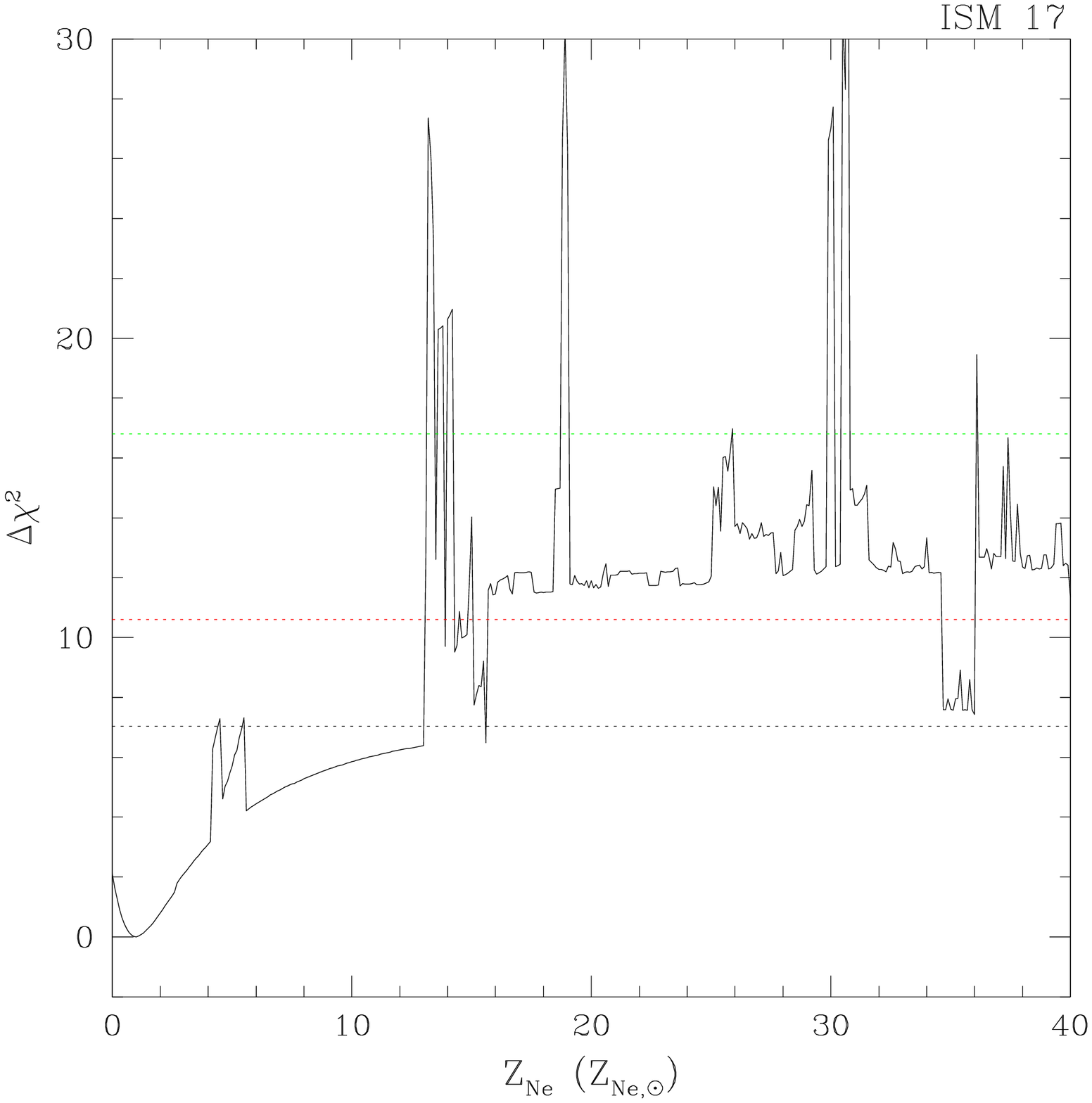}
\caption{Distribution of the $\Delta\chi^2$ values as a function
of the Neon abundance in four of our regions chosen for spectral analysis.
The black dotted line, the red dotted line, and the green dotted line
represents the 68\%, 90\% and 99\% confidence levels, respectively.
\label{chi2ne}}
\end{figure}

\begin{figure}[h]
\epsscale{1.10}
\plottwo{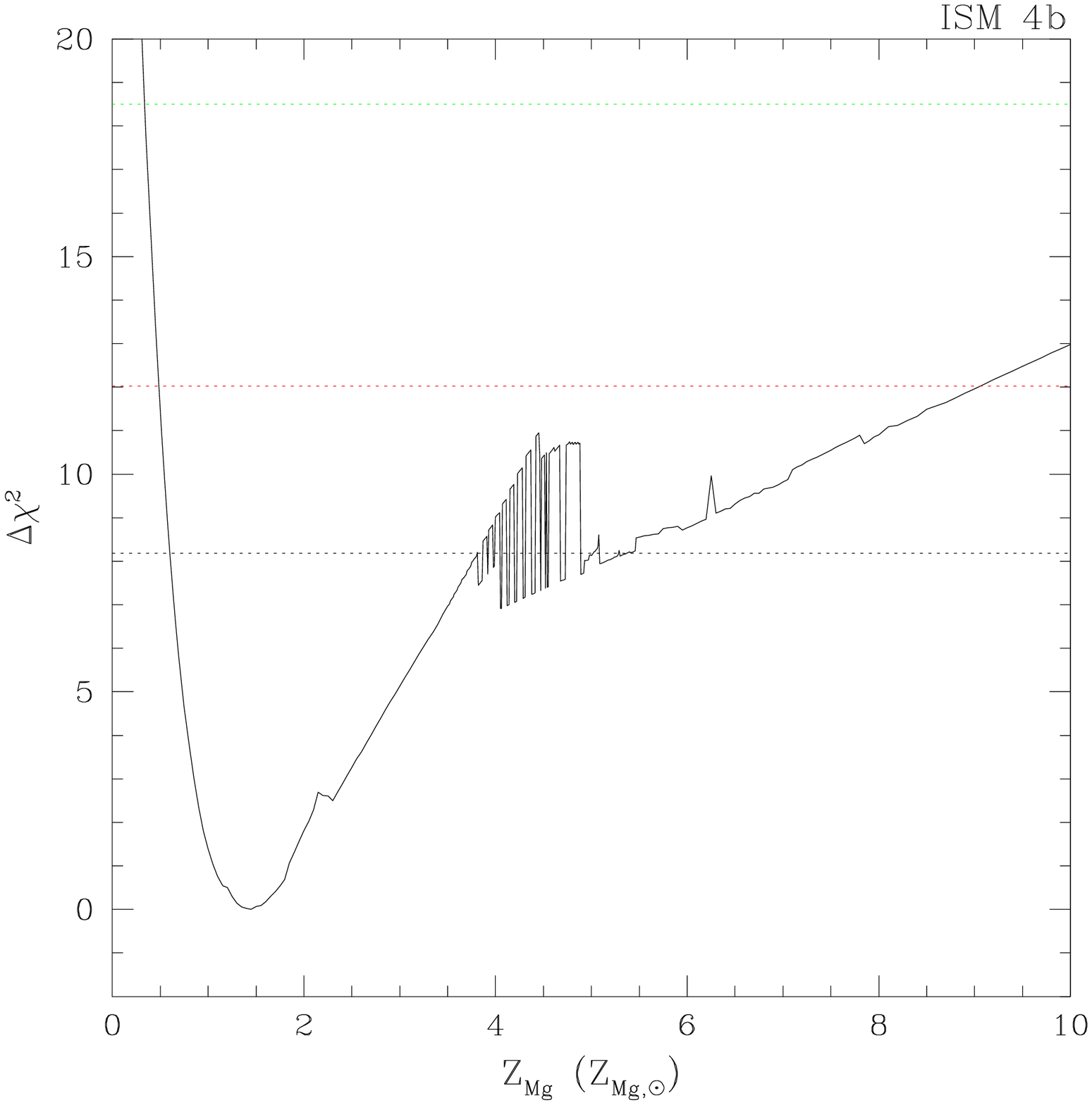}{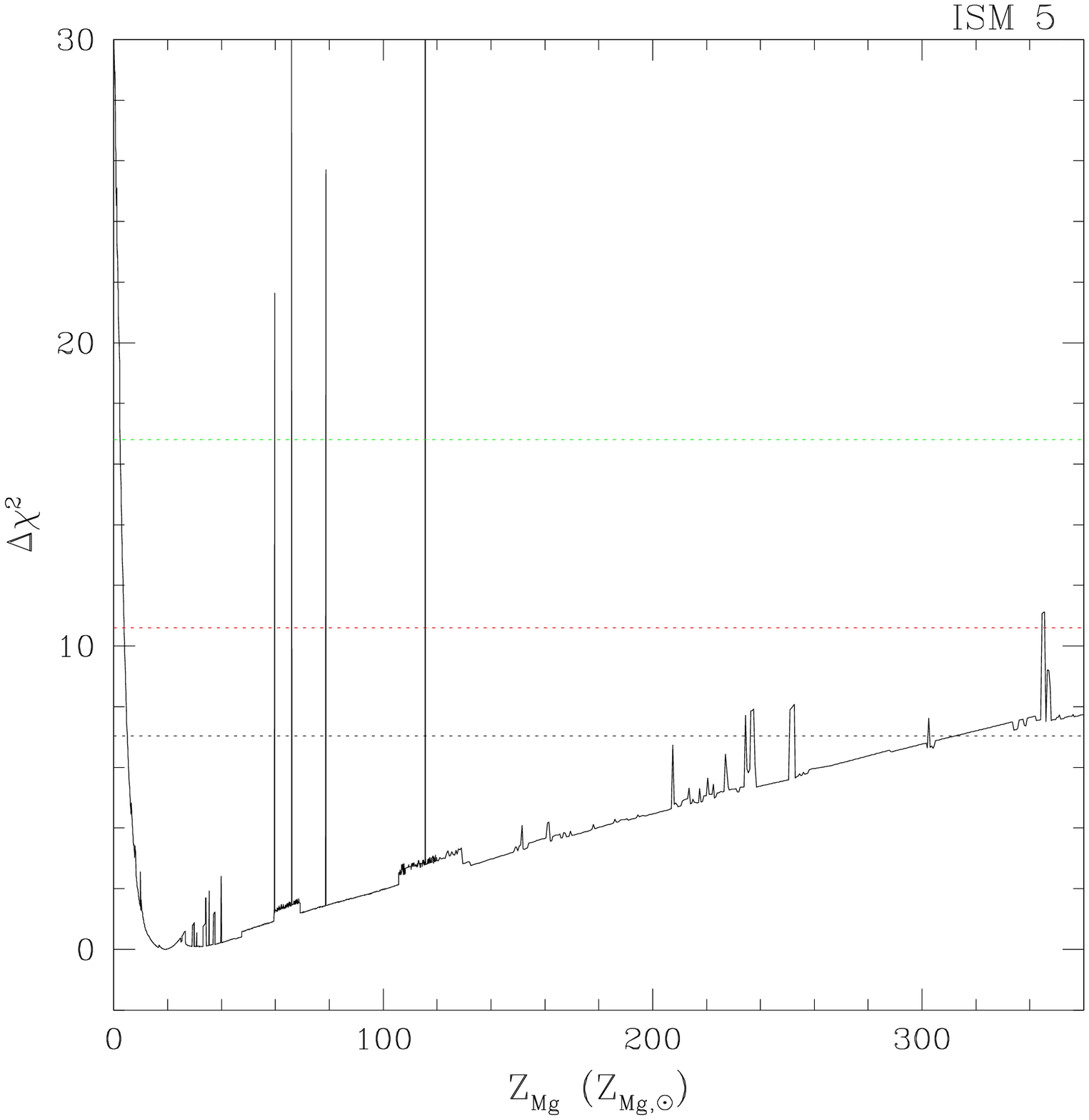}
\plottwo{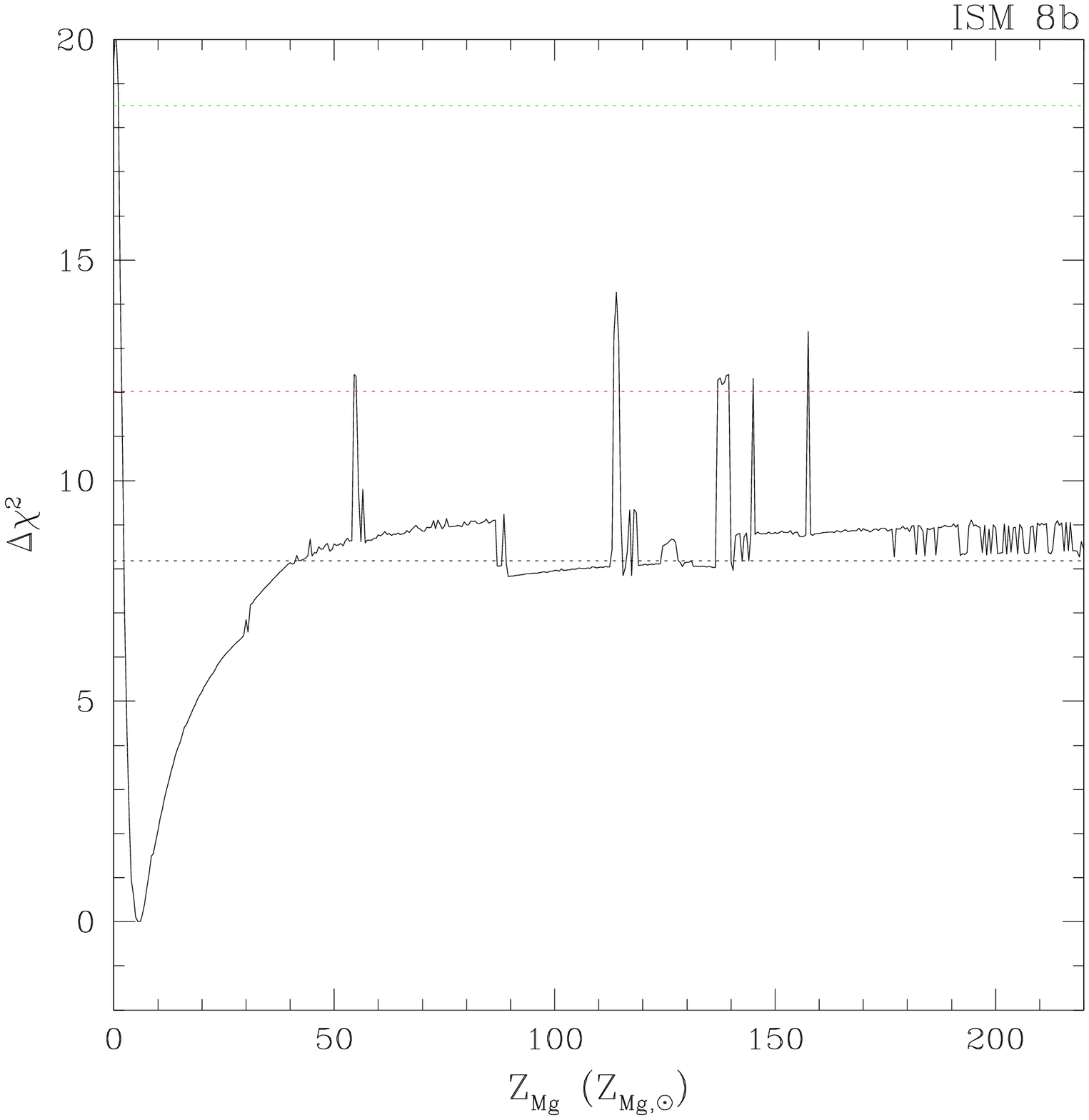}{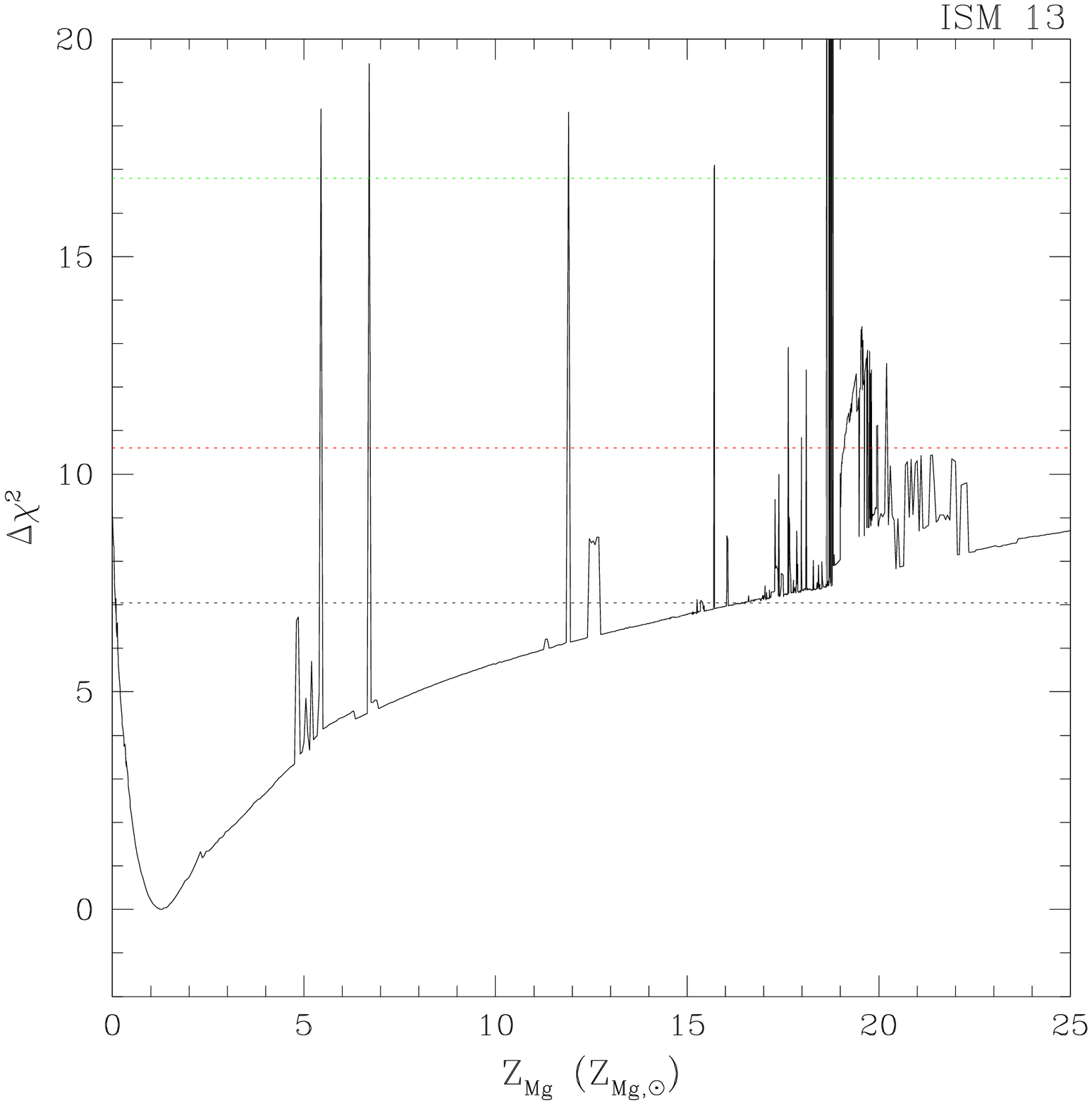}
\caption{Distribution of the $\Delta\chi^2$ values as a function
of the Magnesium abundance in four of our regions chosen for spectral analysis.
The black dotted line, the red dotted line, and the green dotted line
represent the 68\%, 90\% and 99\% confidence levels, respectively.
\label{chi2mg}}
\end{figure}

\begin{figure}[h]
\epsscale{1.10}
\plottwo{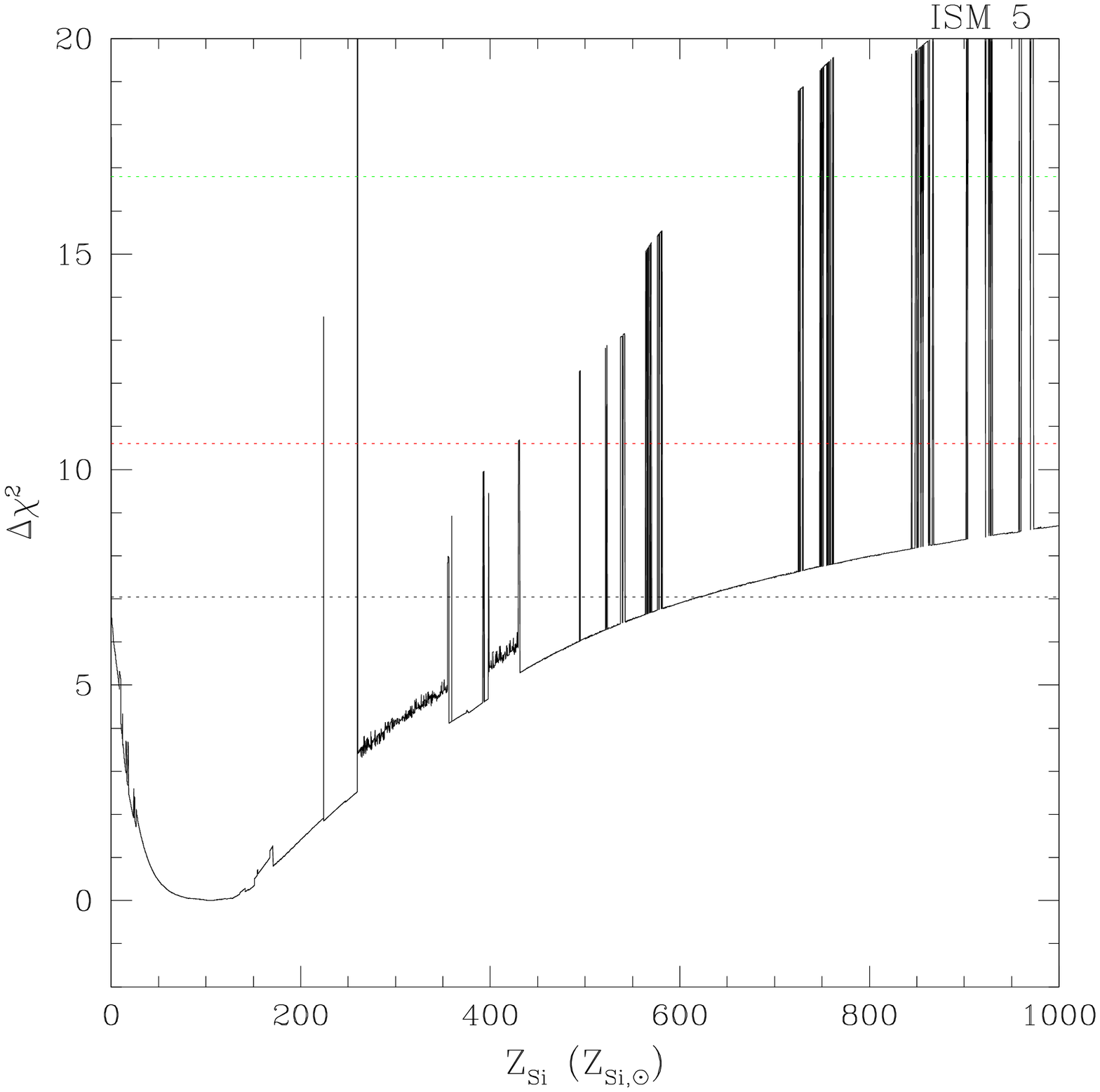}{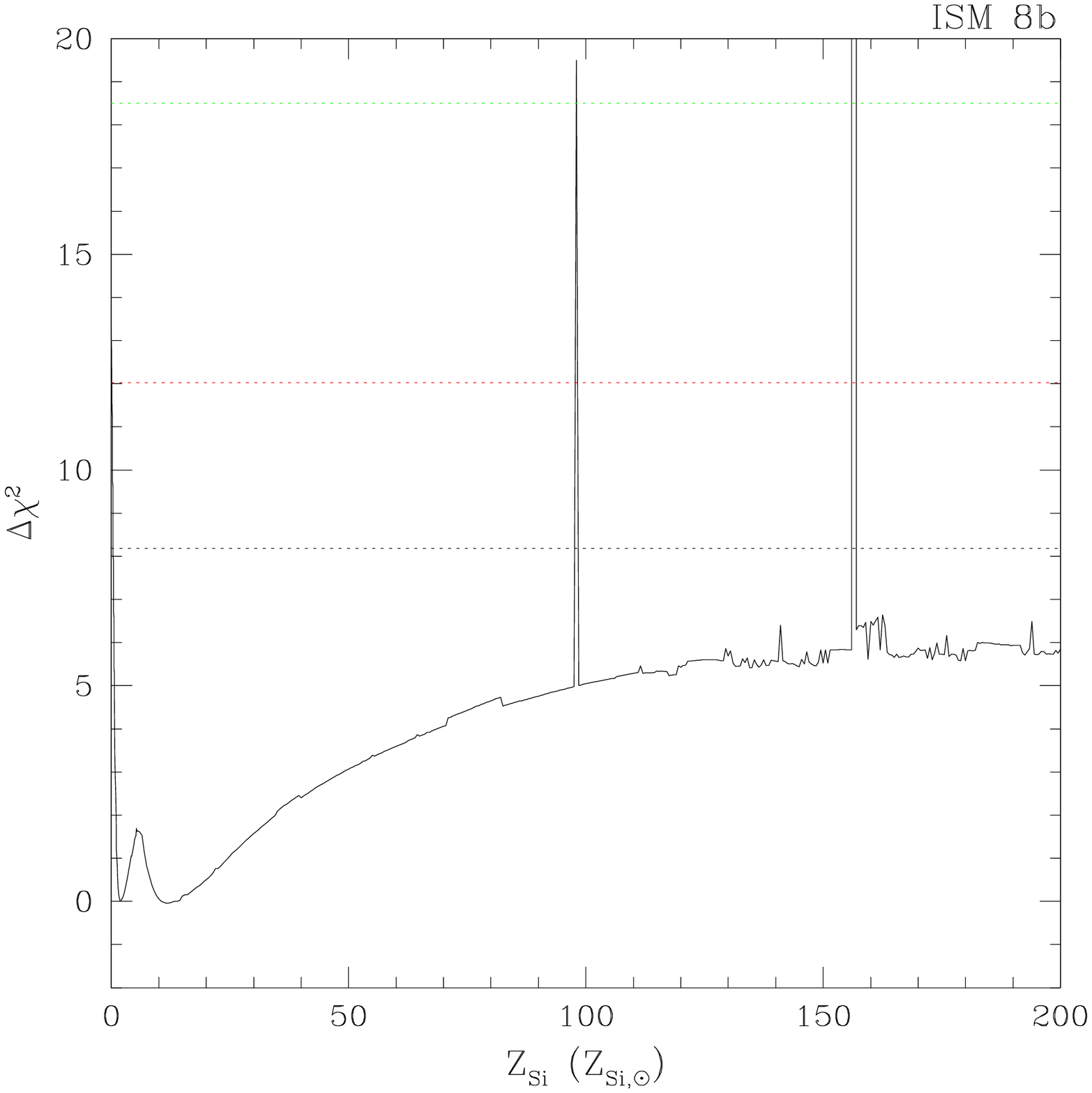}
\plottwo{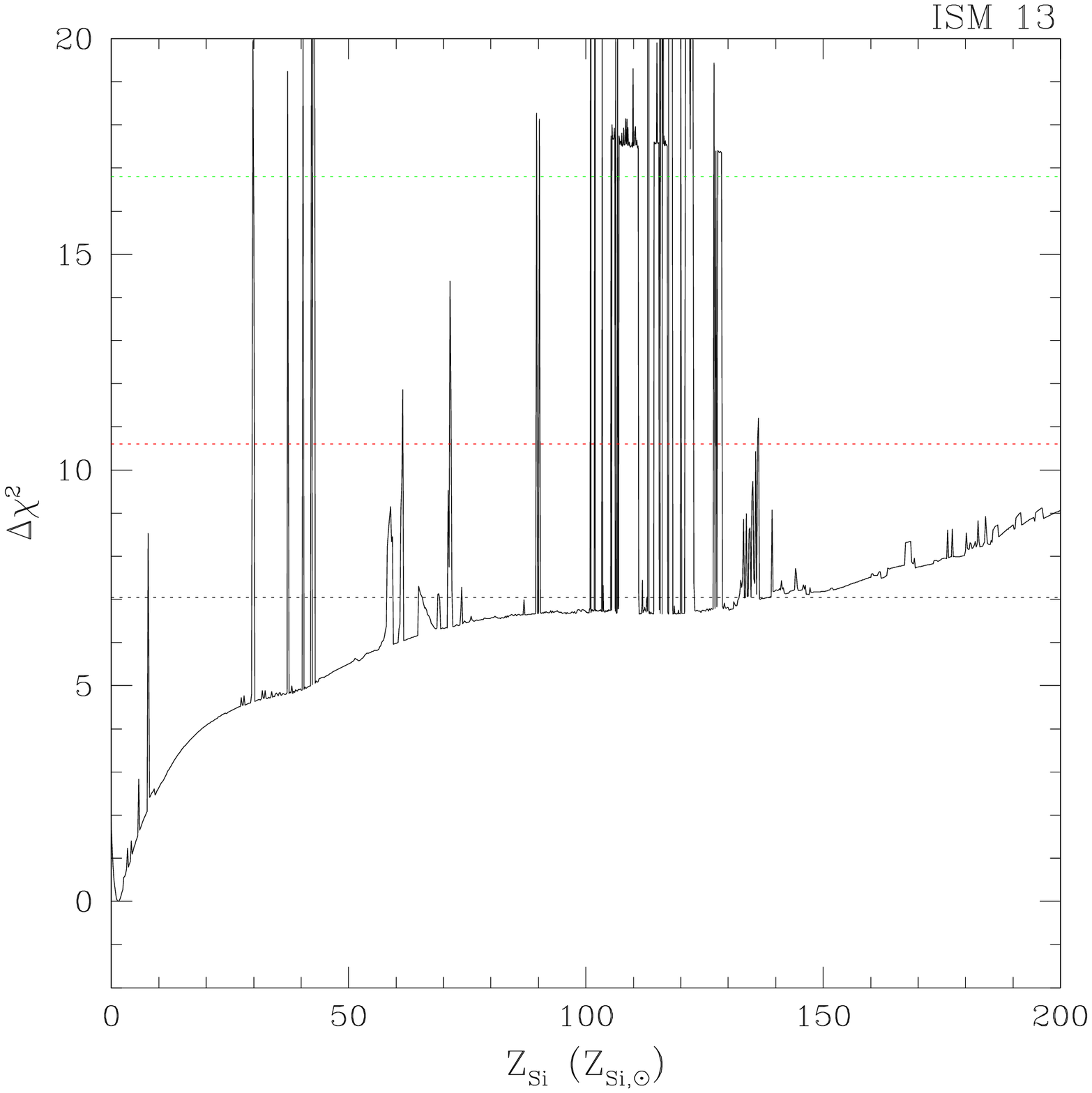}{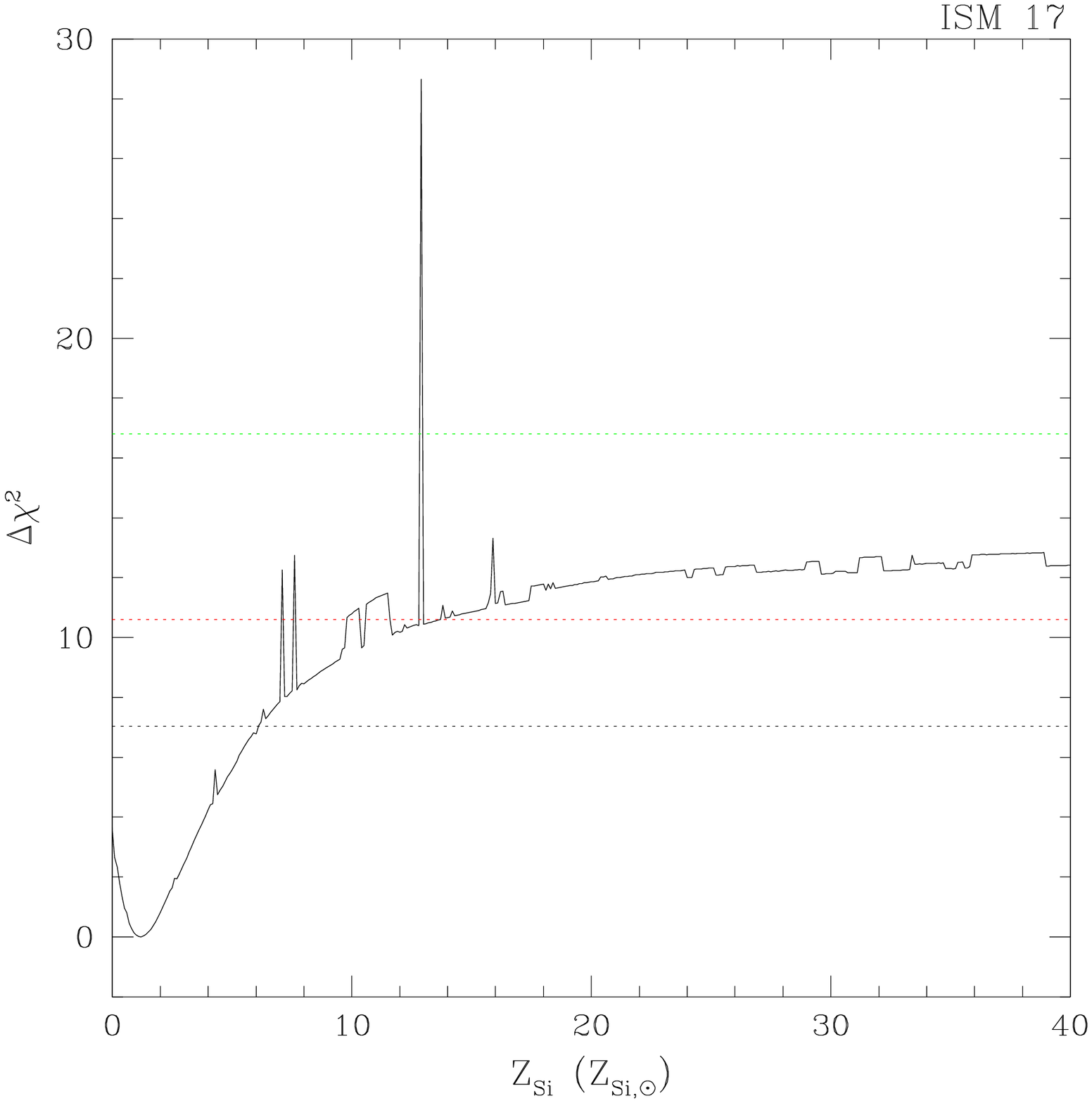}
\caption{Distribution of the $\Delta\chi^2$ values as a function
of the Silicon abundance in four of our regions chosen for spectral analysis.
The black dotted line, the red dotted line, and the green dotted line
represent the 68\%, 90\% and 99\% confidence levels, respectively.
\label{chi2si}}
\end{figure}

\begin{figure}[h]
\epsscale{1.10}
\plottwo{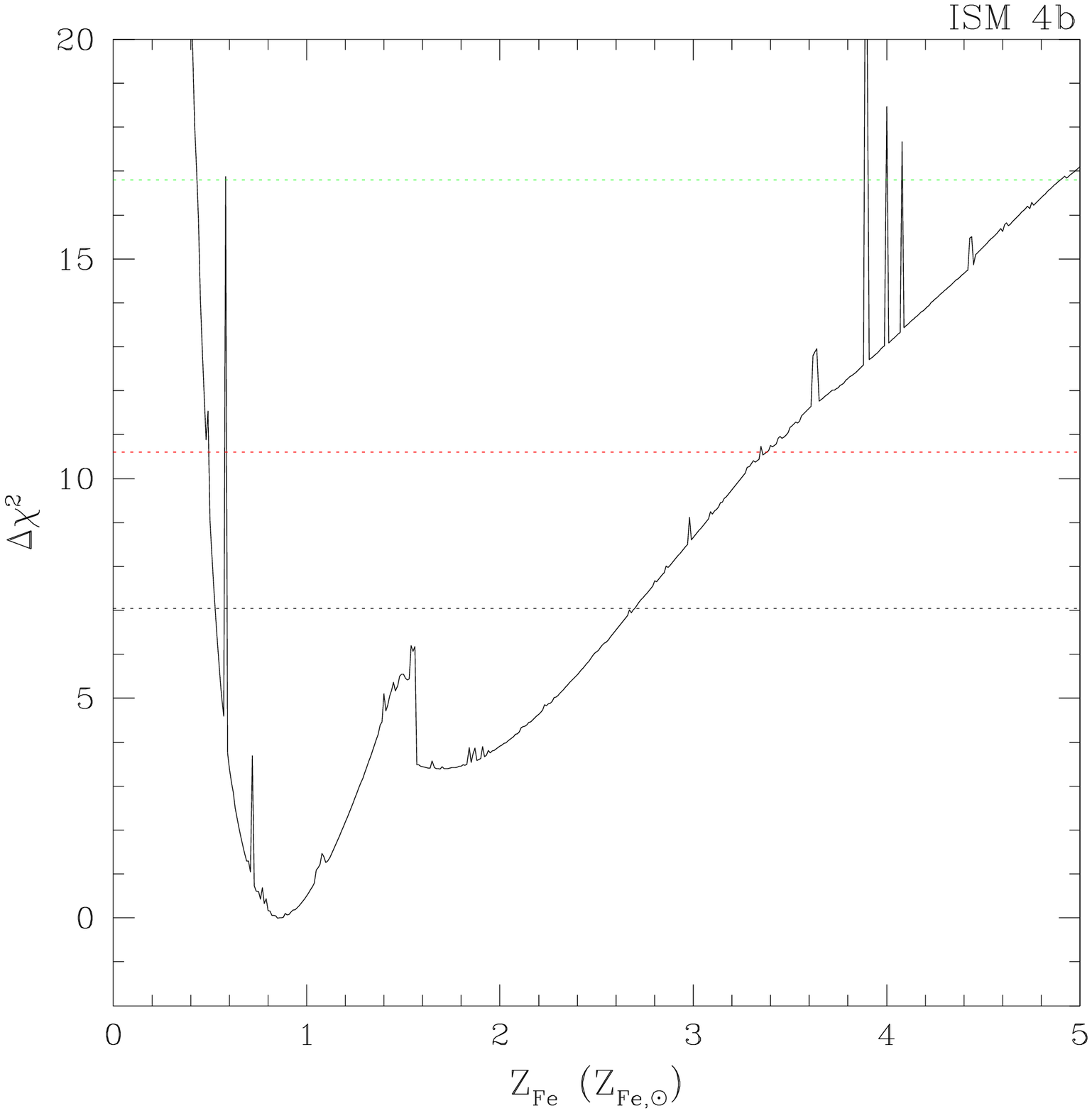}{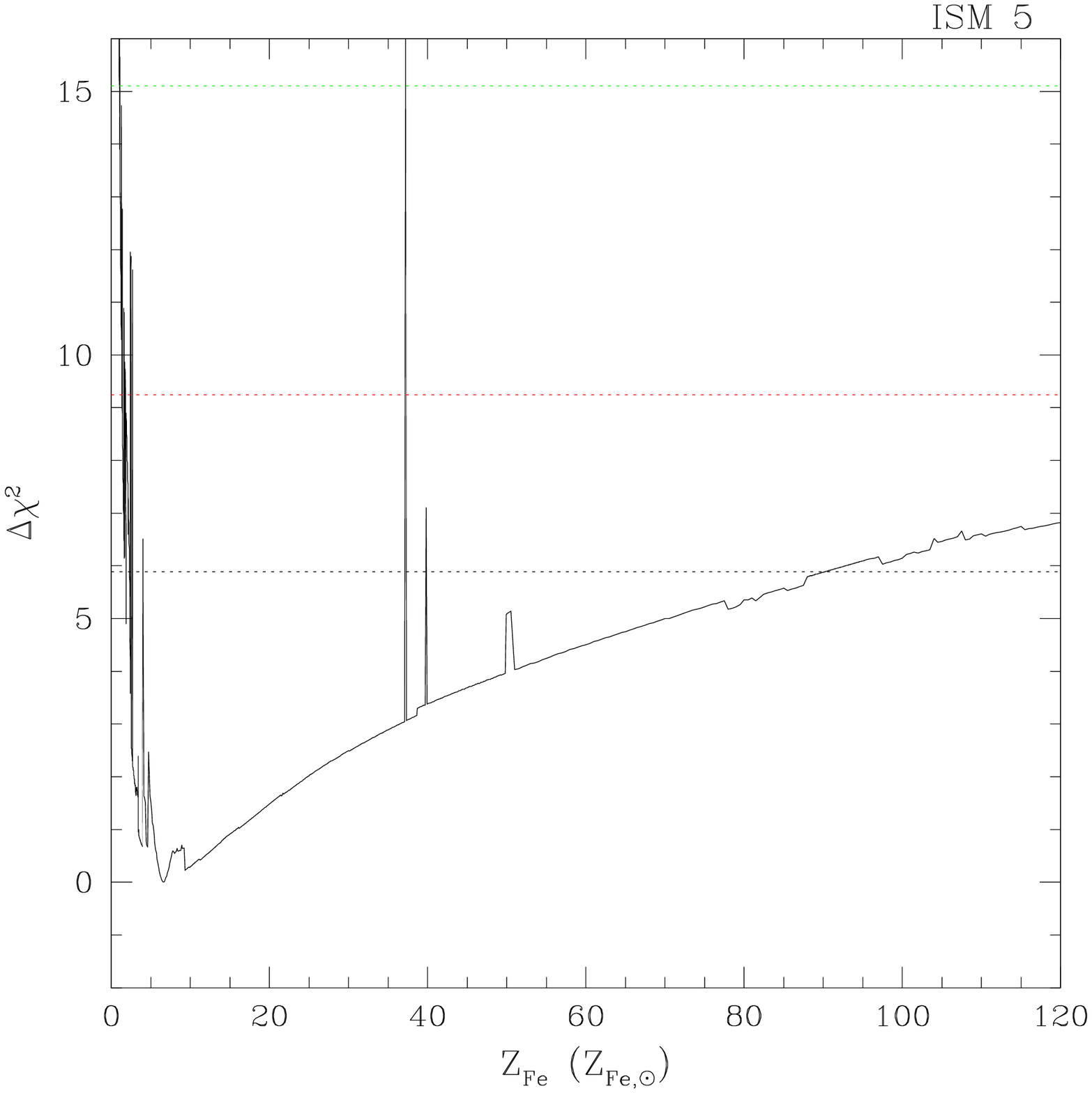}
\plottwo{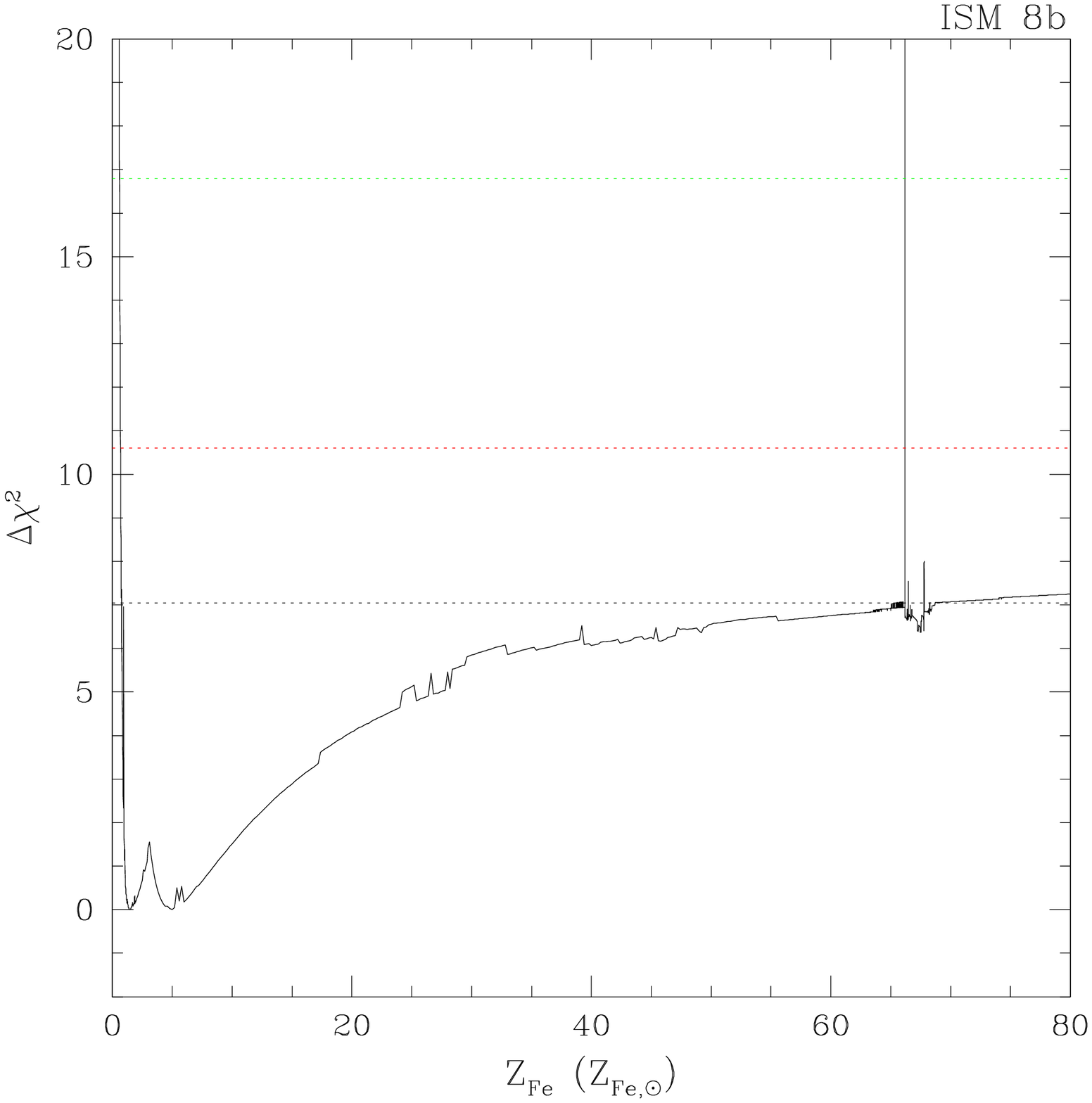}{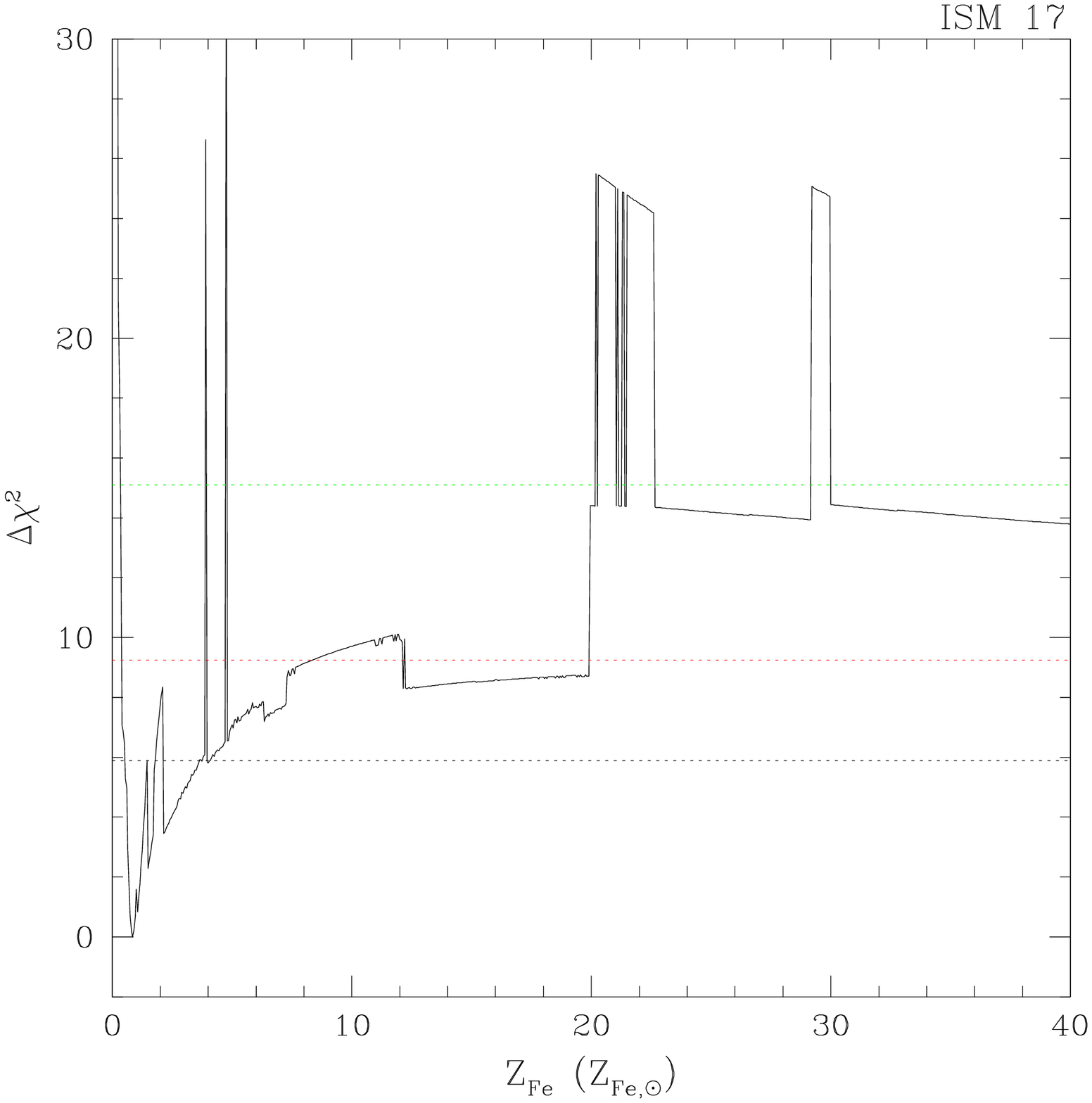}
\caption{Distribution of the $\Delta\chi^2$ values as a function
of the Iron abundance in four of our regions chosen for spectral analysis.
The black dotted line, the red dotted line, and the green dotted line
represent the 68\%, 90\% and 99\% confidence levels, respectively.
\label{chi2fe}}
\end{figure}

The $\chi^2$ parameter space is often asymmetric. 
Therefore, we performed an analysis of the behaviour of $\chi^2$ with each
free parameter (using the {\em steppar} command in XSPEC),
to explore the presence of possible secondary $\chi^2$ minima in the 
confidence range estimation.
Figures~\ref{chi2kt}--\ref{chi2fe} show some examples of the most complicated
features observed in the $\chi^2$ parameter space when varying the temperature ($kT$) 
and the abundances of Ne, Mg, Si and Fe, respectively.
As can be seen in Figure~\ref{chi2kt}, one of the temperatures of Region 4b is
unconstrained even at the $1\sigma$ level. Three other regions (8b, 14 and 17)
show two minima in $\chi^2$ space. For them we adopted $1\sigma$ confidence ranges 
encompassing both minima even when the $\Delta\chi^2$ value between them exceeded the 
$1\sigma$ confidence level, as happens in Regions 8b and 14.
For abundances the presence of secondary minima is more rare: we found a secondary minimum
only in Region 8b (Si and Fe; Figs.~\ref{chi2si} and \ref{chi2fe}) and in Region 4b 
(Fe; Fig.~\ref{chi2fe}).

\subsubsection{Emission-line equivalent width}\label{eqwidth}

\begin{figure}[h]
\epsscale{1.10}
\plottwo{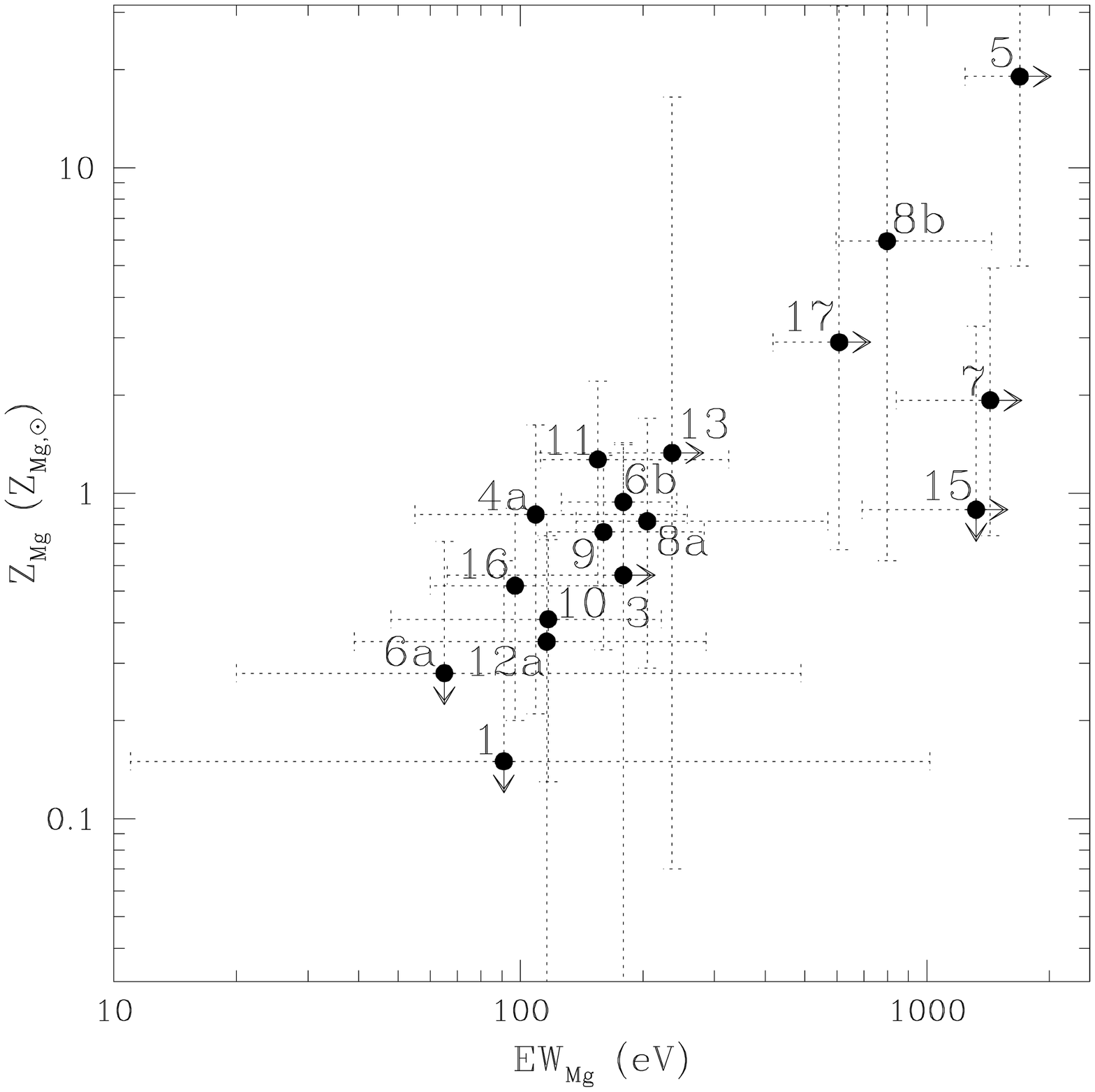}{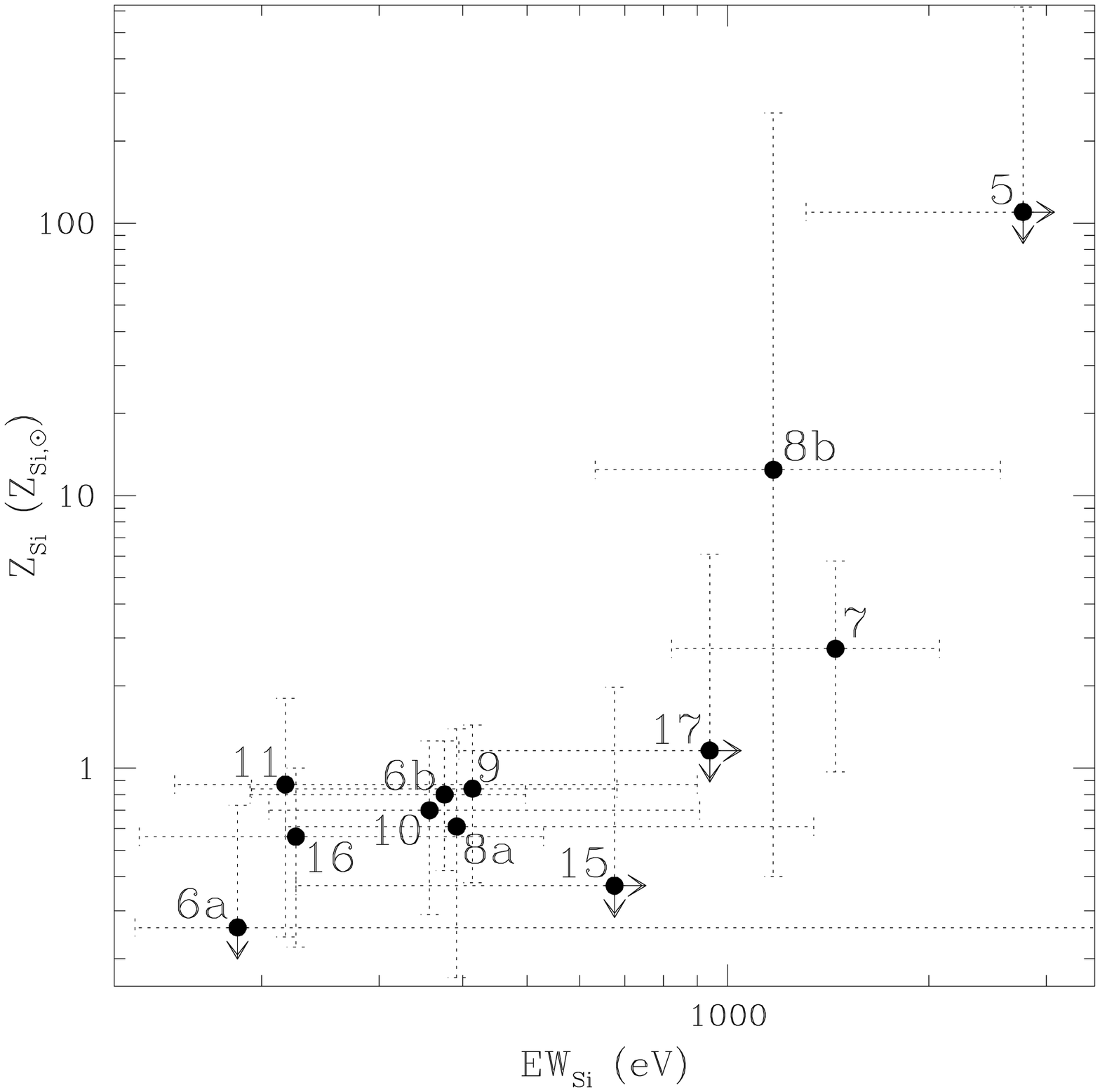}
\caption{Correlation between the measured $EW$ and the abundance measured from the 
spectral fits for Mg (left) and Si (right).
\label{ewvsz}}
\end{figure}

While spectral fits are the standard way to obtain elemental
abundances from X-ray data, we also wish to obtain abundance
limits that are relatively model-independent and that allow us to evaluate
the possible effects of departures from ionization equilibrium (NEI).
Since in our spectra the lines of Mg and Si are not confused with other elements,
they allow us to directly
measure their equivalent widths (hereafter $EW$). For each region we 
considered the spectrum in the whole
0.3--6 keV range and used the best-fit model (as defined in \S~\ref{specfit})
for the continuum. 
We artificially set to zero the abundance of the element in consideration,
adding a gaussian line to the model. Both the normalization and the energy of the gaussian line
were left free to vary in the fit. From the best fit so obtained we measured the
$EW$ of the line using the XSPEC task {\em eqwidth}, using only the thermal component(s) as 
continuum emission.
With {\em eqwidth} we performed a Monte Carlo sampling of the parameter values 
distribution, estimating the errors on the $EW$ (at $1\sigma$) independently
from the $\chi^2$ statistics.
The relation between the measured $EW$ and the measured abundance for Mg and Si are shown in
Figure~\ref{ewvsz}. We find a clear correlation between the two parameters for Mg; a weaker
correlation is suggested by the Si plot, however we must remember that the continuum is much less
well defined for this line, increasing the uncertainties.

The metal abundances measured from the spectral fits can be directly related to the 
$EW$s by the expression:
$$
EW=\frac{a\cdot Z}{1+\alpha\cdot Z},
$$
where $Z$ is the abundance of the element and $a$ and $\alpha$ are two
parameters 
essentially dependent only on the temperature of a hot X-ray emitting gas in a static ionization 
equilibrium. The denominator comes about because the
continuum is produced by bremsstrahlung and the recombination
of lower atomic number elements, while the numerator is simply
proportional to the abundance of the element in question. The
temperature dependence arises mostly from the ionization fractions
of the H-like and He-like ions that produce the emission lines.
The term $\alpha$Z in the denominator accounts for the
contributions
of all the elements other than H and He to the continuum through
bremsstrahlung, recombination, and two-photon emission assuming that
all the elements scale together.  Thus at very high Z the EW will
reach an asymptotic limit, which can only be exceeded if the element
in question is preferentially enhanced relative to the elements that
contribute to the nearby continuum or if NEI increases the equivalent
width. 
In order to assess the likely effects of NEI we computed
equivalent widths from shock wave models using an updated version
of the code described in Raymond (1979) and Cox and Raymond (1985).
While the EW can be far from the equilibrium values just behind the shock,
it tends to be dominated by emission from the region that is not
too far from equilibrium.  We computed models for shocks in the
range 400--900 km/s, corresponding to a temperature immediately behind the shock 
between 0.19 and 0.95 keV. Note that in the latter case the temperature that 
would be derived from the X-ray continuum would be somewhat lower because of contributions
from gas in the cooling region.

\begin{figure}[h]
\epsscale{1.10}
\plottwo{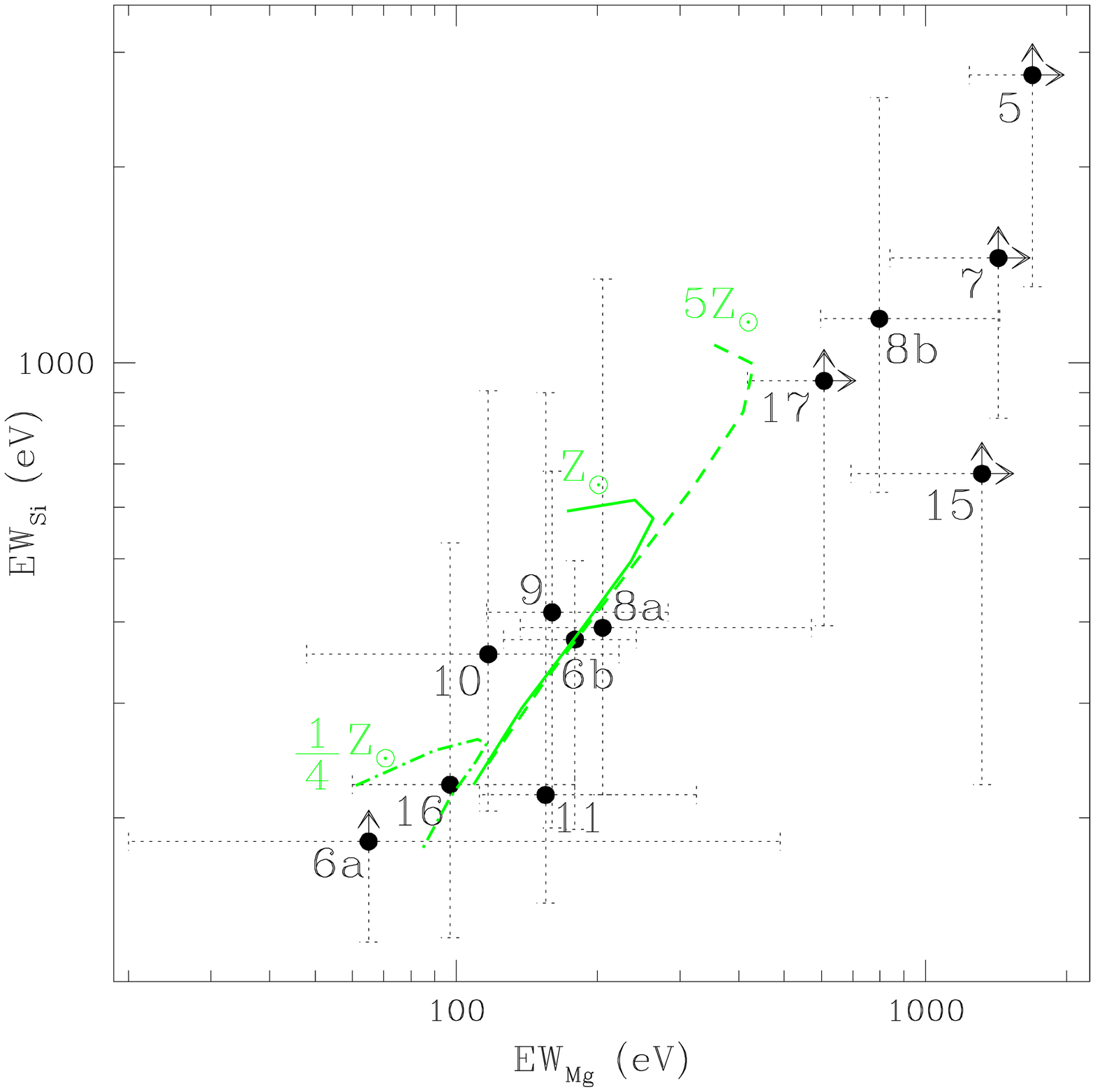}{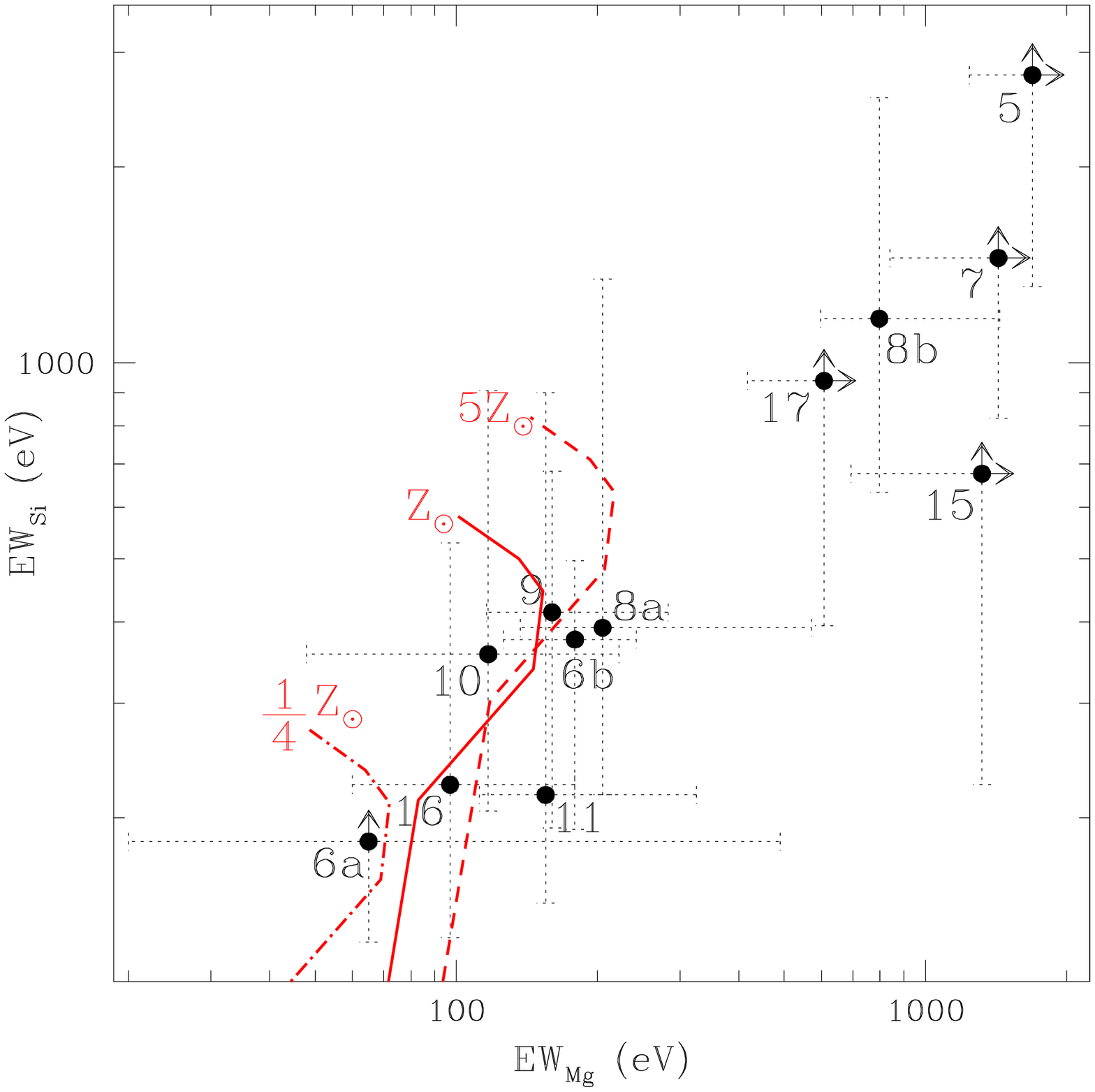}
\caption{Measured $EW$ of the Mg line vs. measured $EW$ of the Si line for the 
hot ISM regions of the spectral analysis. {\it Right:} the green lines represent
the expected $EW$s as a function of the temperature (growing with growing values of
$EW_{Si}$; for details, see text) for a plasma in ionization equilibrium with $Z=\frac{1}{4}Z_\odot$,
$Z=Z_\odot$ and $Z=5Z_\odot$. {\it Right:} the red lines represent
the expected $EW$s as a function of the shock velocity (growing with growing values of
$EW_{Si}$; for details, see text) for a shock-heated plasma with $Z=\frac{1}{4}Z_\odot$,
$Z=Z_\odot$ and $Z=5Z_\odot$
\label{ewmgewsi}}
\end{figure}

Using the values of $\alpha$ and $a$ computed for the two different cases examined (thermal plasma and
shock waves) we derived the
expected $EW$ corresponding to an abundance of $\frac{1}{4}Z_\odot$, $Z_\odot$ and $5Z_\odot$
for both Mg and Si. 
Figure~\ref{ewmgewsi} shows the relation between the $EW$ of Mg and the
$EW$ of Si for all the regions where we were able to obtain a meaningful measure of both.
The regions of the diagram relative to abundances of $\frac{1}{4}Z_\odot$, $Z_\odot$ and
$5Z_\odot$, as a function of the temperature are also plotted.
If we consider the simple equilibrium thermal model (Fig.~\ref{ewmgewsi}, left) the majority of the
points are located in the region corresponding to solar abundance, but there are regions
located in the upper-right portion of the diagram as well. Region 5 has $EW$ not consistent with
$Z\le5Z_\odot$ for both elements, in agreement with the results of the spectral analysis
where this region shows the highest abundances of Mg and Si among all the spectra analyzed.
Also Regions 7, 8b, 15, and 17 show high values of the $EW$, but with large errors, consistently
with what we observe in the abundances measured from the spectral analysis of these regions. 
The situation is even more
dramatic if we consider the shock model (Fig.~\ref{ewmgewsi}, right), where the values of Mg for
the regions listed above are clearly not consistent with
$Z<5Z_\odot$ (as the spectral analysis would suggest), and the Silicon $EW$ for Region 7 is also
not consistent with $Z<5Z_\odot$.
Notice that the values of the $EW$ we derived from the abundances at the different temperatures
were computed assuming that all the elements scale together at the same time.
Thus the continuum due to lower $Z$ elements contributes at the energies of the emission lines. 
If Mg or Si are enhanced relative to O in particular, the equivalent width expected could be
higher.  
That would require that the enrichment in The
Antennae is done by supernovae with a different mass distribution
than those that enriched the solar neighborhood (e.g. Woosley \&
Weaver 1995).
This is indeed likely for very young star forming regions, in which only the most massive stars
will have
completed their evolution.  We note that departures from ionization equilibrium other than those 
of the shock models are possible if the plasma is rapidly heated or cooled. In general they do 
not raise the EW very much 
above the equilibrium values except in a very strongly overionized, cool plasma 
(e.g. Breitschwerdt \& Schmutzler 1994).  While it
is possible to produce such a plasma by rapid adiabatic expansion, the low emissivity makes 
it difficult for such a plasma to make a significant contribution to the spectrum.\\

\subsection{Comparison with results from the first Chandra observation}

Almost all the regions (20 out of 21) did not require the presence of a second
thermal component in the model to fit the data, and half of them (11 out of 21) did
not need the presence of a power-law component to fit the data.  In the cases where
the power-law component was needed, the fixed $\Gamma=1.88$ models were found
to yield a better fit than a thawed $\Gamma$ model in eight regions out of 10.
Only Region 4b required two thermal components with different $kT$. 
In this region, we could not constrain the low $kT$ value (although we found
a minimum in the $\chi^2$ statistics at $kT=0.20$) and we could place only a lower
limit ($kT>0.54$) on the higher temperature component. 
Almost all the single-temperature regions required a temperature of $kT\sim0.6$ keV
to fit the data, with regions 5, 8b, and 14 ($kT\sim0.3$ keV) the only exceptions to this rule.
Considering all the regions analyzed, the temperatures (both in the single- and 
double-temperature cases) range from $\sim0.2$ to $\sim0.8$ keV (considering also
the errors), a range fully consistent with the values found in F03.\\

F03, however, could not reach any conclusions about the metal abundances of the 
ISM because of the shorter integration time of their data set (only ObsID 315).
In contrast, our analysis yields metal abundances that have acceptable constraints in the 
majority of the regions. This 
represents the most striking new result of our spectral
analysis: the abundances of Silicon and Magnesium cover a wide range of values 
from the very low values found in regions like Region 2, 6a, 12a, and 12b 
($Z\sim 0.2 Z_\odot$) to the huge values observed in Region 5 ($Z\ge 20 Z_\odot$).

The intrinsic absorption throughout the hot ISM is generally low, with typical
$N_H\sim10^{20}$ cm$^{-2}$ and often consistent with zero.
The exceptions are the southern nucleus (regions 8a and 8b) and Region 7 (the
dark-blue area in Figure~\ref{truecolor}), which is significantly obscured also in
the optical. Region 7 corresponds to the Overlap Region,
where the most active star formation is now occurring (Mirabel et al.\ 1998; 
Wilson et al.\ 2000; Zhang et al.\ 2001). These regions show
intrinsic absorption corresponding to $N_H \sim1-2\times10^{21}$ cm$^{-2}$.

A direct comparison with the results of the spectral analysis performed on 
ObsID 315 in F03 can be done for the four emission regions identified
and analyzed individually in that paper: the northern nucleus (roughly 
corresponding to our regions 14 and 15),
the southern nucleus (regions 8a, 8b, and 9), the hot spot R1 (regions 12a 
and 12b), and the
hot spot R2 (Region 13).

In the northern nucleus F03 found $kT\sim0.6$ keV, which is fully consistent 
with the temperatures we found in regions 14 and 15 ($0.37_{-0.08}^{+0.39}$ and 
$0.59_{-0.14}^{+0.09}$ keV, respectively). The best-fit abundances given
in F03 ($\sim$0.15 solar) are lower than our solar or slightly sub-solar 
abundances observed in Region 15 (in Region 14 abundances are unconstrained); however,
the large errors encompass our results.

The southern nucleus in F03 is well fitted by a two-temperature model, with temperatures
(0.35 and 0.82 keV) consistent within the errors with the two peaks in $\chi^2$-space observed
for the temperatures of our Region 8b (0.33 and 0.60 keV). However,
in regions 8a and 9 we find a single temperature gas at $\sim$0.6 keV. Also in 
this case the abundances
in F03 are very low ($\sim$0.2 solar), in contrast to abundances consistent with solar
in all the three regions we examined. In this case as well, though, the 
abundances determined by F03 are poorly constrained and consistent with our results.\\

In F03 the hot-spot R1 was fitted with a single-temperature model at $\sim$0.4 keV, 
while the temperatures we find are somewhat higher, with $0.60_{-0.09}^{+0.05}$ keV
and $0.61\pm0.06$ keV for regions 12a and 12b, respectively. F03 determined a value of
$\sim$0.2 solar for the abundance, which is fully consistent with the low values we
find in both Region 12a and Region 12b.

For the hot spot R2 the only comparison possible is for the temperature (F03 
could not constrain the abundance), which was $\sim$0.4 keV in F03, somewhat lower 
than the value in Region 13 ($kT=0.58_{-0.35}^{+0.09}$ keV).  However, the large
errors in our new temperature determination (due to a secondary minimum in the $\chi^2$, 
as described in \S~\ref{errorcomp}) make these two results consistent.

The lower metal-abundance values  suggested by the F03
analysis could be the result of blending of complex emission regions.
Note, however, that in F03 all elements were kept locked at 
the solar ratio, while we have here fitted the abundances of Fe, Mg, Ne, and Si 
individually.

Some of the 21 regions we have chosen are similar to, though
not quite the same as, several of those analyzed by Metz et al.
(2004).  Our regions 2, 4, 5, 6, 12, 13 14 and 15 roughly
correspond to their regions D03, D06, D07, D08, D01, D02, D04
and D05, respectively.  In general, our and their spectral results are
similar, except that the higher quality of our spectra permits
at least a coarse measurement of elemental abundances in contrast
with their assumption of solar abundances throughout the hot ISM.

\subsection{Comparison with {\it ASCA} Results}\label{compare}

The high metal abundances suggested by our spectral analysis in some
regions of The Antennae are at odds with a previous claim, based on {\it ASCA} 
CCD spectra, of an overall extremely low abundance of heavy elements in this 
system ($\sim$0.1
the solar value, Sansom et al.\ 1996). Since the spectral resolution of {\it ASCA}
was comparable to that of {\it Chandra} ACIS, we explore the possibility 
that---besides the lower signal-to-noise ratio of the previous data---poor spatial
resolution may have played a 
role in the low abundances reported from the analysis of the {\it ASCA} observations. 

Given the large ($\sim2$ arcmin) {\it ASCA} beam, the entire emission from The Antennae was 
used to derive a spectrum. To make a fair comparison with the {\it ASCA} data, we 
extracted from our {\it Chandra} data a spectrum of all
21 regions together, leaving the point sources (which could not be excluded in the
{\it ASCA} observation) in the extraction region. We used the procedure described
in \S~\ref{selreg} to extract the spectra and compute response matrices
individually for the first (December 1999) and the
fifth (July 2001) observation, and from a merged event file for the other
five observations combined.  The background was taken from the same areas
used in the case of the individual regions' analysis. The combined spectra and 
response matrices
were used in XSPEC. The spectrum was fitted using a model with two additive 
components:
a Raymond-Smith thermal model (for the diffuse emission) and a power law
(for the point
sources). To repeat the {\it ASCA} analysis we assumed a solar abundance ratio
between the various elements, and fitted the metal abundance with just one parameter.
We considered two different 
components for the absorption, one fixed at the Galactic value 
and the other left free to vary to measure the intrinsic absorption within
The Antennae.
The quality of the fit we obtained is very poor ($\chi^2_{\nu}\sim2.8$), not 
surprisingly
given the complexity  of the temperature and metallicity distributions
throughout The Antennae. However, the best-fit model has a temperature of 
$kT=0.59\pm0.03$ keV
and a metallicity of $0.18_{-0.08}^{+0.15}$ times the solar value, fully
consistent with the {\it ASCA} results (Sansom et al.\ 1996). This demonstrates that the 
analysis of separate ``clean'' regions, possible because of the sub-arcsecond 
resolution
of {\it Chandra}, is needed to detect the metal lines in a complex hot ISM
with CCD spectral resolution.

The above conclusion is reinforced by the intercomparison of the spectral-fit 
results for different subsections of morphologically selected regions (e.g.,
regions 5 and 12b), where the 
line-strength map of Figure~\ref{regions21}a suggested different metal contents in the 
spectra. The 
results of these fits are compiled in Table~\ref{total} and clearly present 
intermediate
metal abundances, typically lower than the peak value. We note that while the 
spatial
resolution of {\it Chandra} surpasses that of any other X-ray telescope it is 
still finite,
and while our exposure of The Antennae is deep the collecting area of 
{\it Chandra} is limited. Therefore, our results on metal abundances by necessity 
still suffer to some degree from an ``averaging'' effect over spectrally complex 
regions. As such, they may still represent lower limits on the local metal enrichment 
in some regions.

This conclusion also agrees with unrelated recent results that demonstrate how 
fitting
thermal plasmas with simplified (single-temperature) emission models may 
underestimate
metal abundances (e.g., Molendi \& Gastaldello 2002; Kim \& Fabbiano 2004; see 
Fabbiano 1995
and Buote \& Fabian 1998 for earlier discussions of this effect).

\section{Summary and Conclusions}

We have performed a detailed study of the X-ray properties of the diffuse 
emission of The Antennae (NGC~4038/39), analyzing the entire 411~ks exposure 
obtained as part of our monitoring observing campaign of this system with 
{\it Chandra} ACIS. Confirming the results of F03, which were based on the first of the seven 
observations used in the present study, we report a spatially and spectrally 
complex hot ISM. Thanks to our deep data, we also detect clear, spatially 
variable emission lines of Ne, Mg, and Si, in addition to the Fe-L blend, which
indicate super-solar metal abundances in a few regions of the hot ISM.

In summary:

(1) We derive an X-ray broad-band color image of the hot ISM, as well as a color  
map displaying the apparent strength of line emission from O+Fe+Ne, Mg-XI, and Si-XIII. 
Both maps demonstrate complexity and have been used to guide our selection of 21 separate 
spectral extraction regions for detailed analysis.

(2) Spectral models combining optically thin thermal-plasma and 
power-law emission can be fitted to our data with the presence, in almost all
regions, of a single thermal 
component with temperatures ranging from 0.2 to 0.7~keV. 
Significant variations of $N_{\rm H}$ throughout the ISM are detected:
column densities range from values 
consistent with the Galactic foreground absorption to higher values of 
$\sim${}$2\times10^{21}$ cm$^{-2}$, as in the southern nucleus and in Region 7, the 
actively star-forming Overlap Region (Mirabel et al.\ 1998; Wilson et 
al.\ 2000; Zhang et al.\ 2001). 
A power-law component is present in half of the regions, showing values
of $\Gamma$ consistent with emission from unresolved X-ray binaries. The study
of this power-law emission will be the subject of a forthcoming paper.

(3) Fitting the Fe-L, Ne-IX, Mg-XI and Si-XIII emission, we find significant 
metal enrichment of the hot ISM. Metal abundances are generally consistent with 
solar, but reach extremes of subsolar and $\sim$20--30 solar in a few 
regions. In supporting evidence, the $EW$ of the emission lines of Mg and Si 
correlates well with the abundances measured.
Previous reports (from ASCA) of subsolar abundances in the hot ISM of the Antennae
can be explained with an "averaging" effect due to the inclusion of the entire
emission (diffuse from different regions and point sources) in a single
spectrum.

The implication of these results for the physical state of the hot ISM and SN
enrichment scenarios are discussed in a companion paper
(Baldi et al., submitted).

\acknowledgements

We thank the {\it Chandra} X-ray Center DS and SDS teams for their efforts in
reducing the data and for developing the software used in the data reduction (SDP)
and analysis (CIAO).
We thank D.-W. Kim for useful discussions.
We also thank the anonymous referee for his careful analysis of the paper
and for giving us helpful suggestions for the presentation of our results.
This work was supported in part by NASA contract NAS8-39073 and NASA grants 
GO1-2115X and GO2-3135X.
F.S. acknowledges partial support from the NSF through grant AST 02-05994.

\clearpage

\clearpage

\begin{figure}[H]
\epsscale{0.95}
\caption{0.3--6 keV image of the central part of The Antennae. The 3$\sigma$
ellipses of the point sources detected by CIAO $wavdetect$ (and reported in
Zezas et al. 2002) are plotted in green.
\label{totimage}}
\end{figure}


\begin{figure}[H]
\epsscale{1.00}
\caption{Adaptively smoothed images of The Antennae after source removal in
the 0.3--0.65 keV band (red), 0.65--1.5 keV band (green) and 1.5--6 keV band 
(blue).
\label{3color}}
\end{figure}


\begin{figure}[H]
\epsscale{1.00}
\caption{Mapped-color smoothed image of X-ray emission in The Antennae. The red
color channel indicates emission from the 0.3--0.65 keV band, green from the
0.65--1.5 keV band, and blue from the 1.5--6 keV band.
\label{truecolor}}
\end{figure}


\begin{figure}[H]
\caption{Color bars representing the possible colors obtainable from the combination of 
red, green and blue channels, referring to the mapped-color smoothed image in Fig.~\ref{truecolor}. 
The bars are labelled with corresponding values of $HR1$ and $HR2$. The
tridimensional cube indicates the cuts in the values of blue counts 
we performed to produce the color bars (see text).
\label{colorbars}}
\end{figure}

\clearpage

\clearpage 

\begin{figure}[H]
\epsscale{0.70}
\epsscale{0.70}
\caption{{\it Top:} Mapped-color line-strength image of The Antennae. Red represents
emission from the Fe-L line blend, green from the Mg line, and blue from the Si line
(for details, see text). The 17 regions used for a preliminary subdivision of 
the hot ISM are marked in white.
{\it Bottom:} 0.3--6 keV image of The Antennae, obtained removing the 3$\sigma$ 
point sources.
The 21 regions used for the spectral analysis of the hot ISM are marked
in white.
\label{regions21}}
\end{figure}


\begin{figure}[h]
\epsscale{1.10}
\caption{{\it Chandra} ACIS-S spectra for regions 1, 2, 3, 4a, 4b, and 5.
Plotted for each region are the spectrum and the data-to-best-fit-model ratios.
\label{spectra1}}
\end{figure}


\begin{figure}[h]
\epsscale{1.10}
\caption{{\it Chandra} ACIS-S spectra for regions 6a, 6b, 7, 8a, 8b, and 9.
Spectrum and data-to-best-fit-model ratios as in Figure~\ref{spectra1}.
\label{spectra2}}
\end{figure}


\begin{figure}[h]
\epsscale{1.10}
\caption{{\it Chandra} ACIS-S spectra for regions 10, 11, 12a, 12b, 13, and 14.
Spectrum and data-to-best-fit-model ratios as in Figure~\ref{spectra1}.
\label{spectra3}}
\end{figure}


\begin{figure}[h]
\epsscale{1.10}
\caption{{\it Chandra} ACIS-S spectra for regions 15, 16, and 17.
Spectrum and data-to-best-fit-model ratios as in Figure~\ref{spectra1}.
\label{spectra4}}
\end{figure}

\clearpage 

\clearpage 

\clearpage 

\clearpage 

\clearpage 

\clearpage

\clearpage 


\begin{deluxetable}{lrrcc}
\tabletypesize{\small}
\tablecaption{Log of {\it Chandra} Observations \label{obslog}}
\tablewidth{0pt}
\tablehead{
\colhead{Instrument} &
\colhead{Obs ID} &
\colhead{Date} &
\colhead{Grating} &
\colhead{Performed duration (ks)}
}
\startdata
ACIS-S&315&December 1, 1999&None&75.53\\
ACIS-S&3040&December 29, 2001&None&63.76\\
ACIS-S&3043&April 18, 2002&None&60.81\\
ACIS-S&3042&May 31, 2002&None&67.27\\
ACIS-S&3044&July 10, 2002&None&36.49\\
ACIS-S&3718&July 13, 2002&None&34.72\\
ACIS-S&3041&November 22, 2002&None&72.91\\
\enddata

\end{deluxetable}


\clearpage 

\begin{deluxetable}{ccccc}
\tablecaption{Net counts in the regions selected for the spectral analysis of
the Diffuse Emission of The Antennae. 
\label{counts}}
\tablewidth{0pt}
\tablehead{
\colhead{Region \#} &
\colhead{cts$_{0.3-0.65keV}$} &
\colhead{cts$_{0.65-1.5keV}$} &
\colhead{cts$_{1.5-6.0keV}$} &
\colhead{cts$_{0.3-6.0keV}$} 
}
\startdata
1 & $205.3\pm16.4$ & $692.6\pm27.2$ & $71.4\pm14.0$ & $969.1\pm34.6$ \\
2 & $162.1\pm13.4$ & $565.8\pm23.9$ & $55.8\pm9.4$ & $783.9\pm29.2$ \\
3 & $218.2\pm17.8$ & $754.3\pm28.8$ & $101.3\pm17.0$ & $1074.1\pm37.9$ \\
4a & $182.9\pm14.5$ & $1114.8\pm33.7$ & $104.3\pm12.5$ & $1402.0\pm38.7$ \\
4b & $374.0\pm20.4$ & $2556.0\pm51.0$ & $176.8\pm16.0$ & $3107.0\pm57.2$ \\
5 & $176.4\pm13.8$ & $876.5\pm29.6$ & $115.0\pm11.9$ & $1167.9\pm35.0$ \\
6a & $374.4\pm20.4$ & $1733.3\pm42.0$ & $231.4\pm17.5$ & $2339.1\pm49.8$ \\
6b & $520.2\pm24.3$ & $3324.3\pm58.0$ & $286.7\pm20.7$ & $4131.7\pm65.8$ \\
7 & $115.8\pm13.7$ & $1236.6\pm35.8$ & $379.1\pm22.7$ & $1731.3\pm44.4$ \\
8a & $407.9\pm20.8$ & $3772.4\pm61.7$ & $600.8\pm25.5$ & $4781.9\pm70.0$ \\
8b & $313.6\pm18.3$ & $3128.4\pm56.0$ & $335.3\pm19.4$ & $3777.3\pm62.1$ \\
9 & $714.4\pm29.2$ & $5094.6\pm74.1$ & $703.3\pm30.9$ & $6510.2\pm82.3$ \\
10 & $583.5\pm28.4$ & $2495.0\pm51.9$ & $271.1\pm26.7$ & $3349.8\pm65.0$ \\
11 & $216.8\pm16.0$ & $1341.1\pm37.0$ & $174.2\pm15.7$ & $1732.1\pm43.2$ \\
12a & $338.6\pm18.8$ & $1308.2\pm36.2$ & $81.7\pm10.6$ & $1728.3\pm42.4$ \\
12b & $153.5\pm12.7$ & $662.1\pm25.9$ & $37.7\pm7.2$ & $853.5\pm29.6$ \\
13 & $119.4\pm11.3$ & $561.3\pm23.9$ & $70.5\pm9.2$ & $751.0\pm28.0$ \\
14 & $53.2\pm7.6$ & $633.3\pm25.1$ & $68.8\pm8.8$ & $755.1\pm28.0$ \\
15 & $131.6\pm11.8$ & $1031.3\pm32.1$ & $163.9\pm13.3$ & $1326.7\pm36.6$ \\
16 & $482.3\pm24.3$ & $2530.8\pm51.0$ & $245.6\pm20.8$ & $3258.4\pm60.1$ \\
17 & $189.7\pm18.6$ & $1296.3\pm37.4$ & $306.4\pm24.7$ & $1792.6\pm48.6$ \\
\enddata

\end{deluxetable}
\clearpage


\clearpage 

\begin{deluxetable}{cccccccccccc}
\tabletypesize{\scriptsize}
\rotate
\tablecaption{Best-Fit Parameters (1$\sigma$ errors for one interesting parameter) 
for the Diffuse Emission of The Antennae. 
\label{mytable1old}}
\tablewidth{0pt}
\tablehead{
\colhead{Region \#} &
\colhead{$\chi^2/dof$} &
\colhead{\begin{tabular}{c}
$N_H$\\
($\times10^{20}$ cm$^{-2}$)
\end{tabular}} &
\colhead{\begin{tabular}{c}
$kT$\\
(keV)
\end{tabular}} &
\colhead{\begin{tabular}{c}
$Z_{Ne}$\\
($\times Z_{Ne,\odot}$)
\end{tabular}} &
\colhead{\begin{tabular}{c}
$Z_{Mg}$\\
($\times Z_{Mg,\odot}$)
\end{tabular}} &
\colhead{\begin{tabular}{c}
$Z_{Si}$\\
($\times Z_{Si,\odot}$)
\end{tabular}} &
\colhead{\begin{tabular}{c}
$Z_{Fe}$\\
($\times Z_{Fe,\odot}$)
\end{tabular}} &
\colhead{$\Gamma$} &
\colhead{Mod\tablenotemark{(1)}}
}
\startdata
1&43.5/38&$<2.68$&$0.61_{-0.03}^{+0.02}$&$<0.28$&$0.23_{-0.18}^{+0.17}$&$<0.19$&$0.20_{-0.04}^{+0.03}$&\nodata&A1\\
\hline
2&34.3/25&$<0.79$&$0.62_{-0.04}^{+0.03}$&$0.26_{-0.26}^{+0.33}$&$0.28_{-0.19}^{+0.21}$&$0.08_{-0.08}^{+0.36}$&$0.22\pm0.04$&\nodata&A1\\
\hline
3&57.8/48&$<1.19$&$0.54\pm0.04$&$0.39_{-0.30}^{+0.27}$&$0.47_{-0.23}^{+0.31}$&$0.38_{-0.34}^{+0.46}$&$0.23_{-0.04}^{+0.07}$&\nodata&A1\\
\hline
4a&51.1/44&$1.42_{-1.42}^{+2.83}$&$0.63\pm0.02$&$0.29_{-0.29}^{+0.34}$&$0.80_{-0.30}^{+0.36}$&$0.27_{-0.18}^{+0.22}$&$0.38_{-0.07}^{+0.08}$&
\nodata&A1\\
\hline
4b&88.2/63&$0.97_{-0.97}^{+1.88}$&\begin{tabular}{c}
$0.20\pm0.04$\\ 
$0.60\pm0.02$
\end{tabular}
&$1.63_{-0.58}^{+0.92}$&$2.02_{-0.59}^{+0.97}$&$1.59_{-0.45}^{+0.75}$&$0.96_{-0.18}^{+0.32}$&\nodata&A2\\
\hline
5&31.4/35&$0.91_{-0.91}^{+1.23}$&$0.30\pm0.02$&$8.02_{-1.88}^{+2.58}$&$17.28_{-4.43}^{+8.09}$&$24.31_{-11.05}^{+15.80}$&
$2.76_{-0.60}^{+0.73}$&$1.88$\tablenotemark{(2)}&B1\tablenotemark{(2)}\\
\hline
6a&74.6/61&$0.74_{-0.74}^{+1.69}$&$0.66\pm0.02$&$0.14_{-0.14}^{+0.19}$&$0.28_{-0.13}^{+0.15}$&$0.24_{-0.12}^{+0.14}$&$0.24\pm0.03$&
$-0.61_{-0.17}^{+0.22}$&B1\\
\hline
6b&95.5/80&$1.47_{-1.12}^{+1.16}$&$0.61\pm0.01$&$0.66_{-0.19}^{+0.21}$&$0.85_{-0.16}^{+0.18}$&$0.70_{-0.15}^{+0.17}$&$0.42_{-0.03}^{+0.04}$&
\nodata&A1\\
\hline
7&86.9/65&$13.77_{-4.05}^{+5.34}$&$0.56_{-0.05}^{+0.04}$&$2.68_{-1.33}^{+3.12}$&$2.93_{-1.43}^{+3.48}$&$3.27_{-1.68}^{+3.31}$&
$0.59_{-0.23}^{+0.50}$&$1.88$\tablenotemark{(2)}&B1\tablenotemark{(2)}\\
\hline
8a&102.5/90&$8.66_{-1.69}^{+1.80}$&$0.61_{-0.02}^{+0.01}$&$1.55_{-0.38}^{+0.47}$&$0.87_{-0.22}^{+0.29}$&$0.86_{-0.20}^{+0.27}$&
$0.58_{-0.08}^{+0.09}$&$1.88$\tablenotemark{(2)}&B1\tablenotemark{(2)}\\
\hline
8b&82.5/71&$24.61_{-1.45}^{+3.82}$&$0.33\pm0.02$&$15.44_{-6.39}^{+9.67}$&$8.51_{-3.95}^{+8.67}$&$17.58_{-8.37}^{+25.61}$&
$8.37_{-3.51}^{+9.79}$&$3.61_{-0.09}^{+0.22}$&B1\\
\hline
9&120.6/108&$2.82_{-1.00}^{+1.18}$&$0.62\pm0.01$&$1.29_{-0.29}^{+0.33}$&$0.90_{-0.19}^{+0.22}$&$0.93_{-0.19}^{+0.22}$&
$0.57\pm0.07$&$1.88$\tablenotemark{(2)}&B1\tablenotemark{(2)}\\
\hline
10&128.2/93&$<0.31$&$0.63\pm0.02$&$0.33\pm0.16$&$0.31_{-0.10}^{+0.11}$&$0.57_{-0.15}^{+0.16}$&$0.27\pm0.02$&\nodata&A1\\
\hline
11&57.2/54&$1.88_{-1.88}^{+2.69}$&$0.62\pm0.02$&$0.64_{-0.33}^{+0.40}$&$1.43_{-0.40}^{+0.31}$&$1.02_{-0.35}^{+0.44}$&$0.38_{-0.06}^{+0.08}$&
\nodata&A1\\
\hline
12a&55.0/46&$<0.34$&$0.60\pm0.02$&$0.15_{-0.15}^{+0.19}$&$0.24_\pm0.12$&$0.08_{-0.08}^{+0.14}$&$0.26\pm0.03$&\nodata&A1\\
\hline
12b&24.5/28&$<0.12$&$0.61\pm0.02$&$<0.12$&$0.20_{-0.17}^{+0.18}$&$<0.14$&$0.26_{-0.03}^{+0.05}$&\nodata&A1\\
\hline
13&32.2/22&$0.32_{-0.32}^{+2.44}$&$0.58_{-0.08}^{+0.04}$&$0.32_{-0.32}^{+0.74}$&$0.80_{-0.38}^{+0.39}$&$0.62_{-0.58}^{+0.67}$&
$0.35_{-0.08}^{+0.09}$&$1.88$\tablenotemark{(2)}&B1\tablenotemark{(2)}\\
\hline
14&28.8/21&$3.07_{-3.07}^{+5.25}$&$0.37/pm0.03$&$1.92_{-1.92}^{2.02}$&$1.11_{-0.89}^{+1.48}$&$1.93_{-1.31}^{+2.08}$&
$0.65_{-0.18}^{+0.37}$&$1.88$\tablenotemark{(2)}&B1\tablenotemark{(2)}\\
\hline
15&44.9/38&$3.95_{-3.52}^{+4.08}$&$0.59\pm0.03$&$1.33_{-0.70}^{+1.24}$&$1.41_{-0.58}^{+1.08}$&$0.65_{-0.57}^{+0.66}$&
$0.66_{-0.18}^{+0.23}$&$1.88$\tablenotemark{(2)}&B1\tablenotemark{(2)}\\
\hline
16&67.0/78&$0.76_{-0.76}^{+1.37}$&$0.62\pm0.01$&$0.31_{-0.17}^{+0.19}$&$0.54_{-0.14}^{+0.16}$&$0.59_{-0.15}^{+0.17}$&$0.33_{-0.03}^{+0.04}$&
\nodata&A1\\
\hline
17&90.5/73&$3.08_{-1.95}^{+2.75}$&$0.62_{-0.02}^{+0.03}$&$1.60_{-0.87}^{+1.56}$&$2.94_{-1.14}^{+2.25}$&$1.39_{-0.75}^{+1.14}$&
$0.80_{-0.21}^{+0.32}$&$1.88$\tablenotemark{(2)}&B1\tablenotemark{(2)}\\
\enddata

\tablenotetext{(1)}{XSPEC Models: A1 = {\em wabs(wabs(vapec))}; A2 = {\em 
wabs(wabs(vapec+vapec))}; 
B1 = {\em wabs(wabs(vapec+powerlaw))}; B2 = 
{\em wabs(wabs(vapec+vapec+powerlaw))}.}
\tablenotetext{(2)}{Photon index frozen.}

\end{deluxetable}
\clearpage



\clearpage 

\begin{deluxetable}{cccccccccc}
\tabletypesize{\scriptsize}
\rotate
\tablecaption{Best-Fit Parameters ($Z_\alpha$ and $Z_{Fe}$ determined as described in
the text and 1$\sigma$ errors computed for all interesting parameters) 
for the Diffuse Emission of The Antennae. 
\label{mytable1}}
\tablewidth{0pt}
\tablehead{
\colhead{Region \#} &
\colhead{\begin{tabular}{c}
$N_H$\\
($\times10^{20}$ cm$^{-2}$)
\end{tabular}} &
\colhead{\begin{tabular}{c}
$kT$\\
(keV)
\end{tabular}} &
\colhead{\begin{tabular}{c}
$Z_{Ne}$\\
($\times Z_{Ne,\odot}$)
\end{tabular}} &
\colhead{\begin{tabular}{c}
$Z_{Mg}$\\
($\times Z_{Mg,\odot}$)
\end{tabular}} &
\colhead{\begin{tabular}{c}
$Z_{Si}$\\
($\times Z_{Si,\odot}$)
\end{tabular}} &
\colhead{\begin{tabular}{c}
$Z_{Fe}$\\
($\times Z_{Fe,\odot}$)
\end{tabular}} &
\colhead{$\Gamma$} 
}
\startdata
1&$<8.60$&$0.61_{-0.10}^{+0.07}$&$<0.75$&$0.15_{-0.15}^{+0.47}$&$<0.51$&$0.20_{-0.08}^{+0.09}$&\nodata\\
\hline
2&$<4.95$&$0.62_{-0.18}^{+0.10}$&$0.20_{-0.20}^{+0.84}$&$0.35_{-0.35}^{+0.61}$&$0.19_{-0.19}^{+2.69}$&$0.20_{-0.10}^{+0.12}$&\nodata\\
\hline
3&$<6.73$&$0.54_{-0.14}^{+0.09}$&$0.48_{-0.48}^{+1.01}$&$0.56_{-0.54}^{+0.85}$&$0.51_{-0.51}^{+1.80}$&$0.26_{-0.11}^{+0.16}$&\nodata\\
\hline
4a&$1.42_{-1.42}^{+8.10}$&$0.63_{-0.06}^{+0.07}$&$0.38_{-0.38}^{+0.78}$&$0.86_{-0.65}^{+0.76}$&$0.23_{-0.23}^{+0.56}$&$0.44_{-0.13}^{+0.14}$&
\nodata\\
\hline
4b&$0.97_{-0.97}^{+4.22}$&\begin{tabular}{c}
$0.20 (unconstr.)$\\ 
$>0.54$
\end{tabular}
&$1.35_{-0.90}^{+1.46}$&$1.44_{-0.84}^{+3.93}$&$1.21_{-0.80}^{+1.13}$&$0.86_{-0.34}^{+1.84}$&\nodata\\
\hline
5&$0.91_{-0.91}^{+10.69}$&$0.30_{-0.08}^{+0.07}$&$8.86_{-4.24}^{+12.47}$&
$19.07_{-14.08}^{+293.43}$&$109.82_{-109.82}^{+510.25}$&$2.81_{-0.95}^{+87.43}$&
$1.88$\tablenotemark{(1)}\\
\hline
6a&$0.74_{-0.74}^{+5.68}$&$0.66_{-0.07}^{+0.06}$&$0.16_{-0.16}^{+0.56}$&$0.28_{-0.28}^{+0.43}$&$0.26_{-0.26}^{+0.47}$&$0.27\pm0.09$&
$-0.61_{-0.39}^{+1.15}$\\
\hline
6b&$1.47_{-1.47}^{+1.79}$&$0.61\pm0.04$&$0.72_{-0.39}^{+0.44}$&$0.94_{-0.37}^{+0.49}$&$0.80_{-0.38}^{+0.46}$&$0.43_{-0.09}^{+0.12}$&\nodata\\
\hline
7&$13.77_{-9.39}^{+11.69}$&$0.56_{-0.16}^{+0.12}$&$2.35_{-1.83}^{+3.85}$&$1.93_{-1.19}^{+3.00}$&$2.74_{-1.77}^{+3.02}$&
$0.46_{-0.18}^{+0.43}$&$1.88$\tablenotemark{(1)}\\
\hline
8a&$8.66_{-4.81}^{+5.94}$&$0.61_{-0.05}^{+0.04}$&$1.50_{-0.87}^{+1.65}$&$0.82_{-0.53}^{+0.88}$&$0.61_{-0.44}^{+0.78}$&
$0.70_{-0.19}^{+0.31}$&$1.88$\tablenotemark{(1)}\\
\hline
8b&$24.61_{-11.43}^{+11.60}$&$0.33_{-0.05}^{+0.28}$&$15.57_{-14.62}^{+\infty}$&$5.95_{-5.33}^{+132.59}$&$12.44_{-12.04}^{+\infty}$&
$5.11_{-4.35}^{+63.59}$&$3.61_{-0.70}^{+0.64}$\\
\hline
9&$2.82_{-2.82}^{+3.86}$&$0.62_{-0.05}^{+0.04}$&$1.26_{-0.67}^{+0.99}$&$0.76_{-0.43}^{+0.55}$&$0.84_{-0.46}^{+0.60}$&
$0.59_{-0.14}^{+0.20}$&$1.88$\tablenotemark{(1)}\\
\hline
10&$<2.04$&$0.63_{-0.05}^{+0.04}$&$0.33_{-0.33}^{+0.41}$&$0.41_{-0.28}^{+0.31}$&$0.70_{-0.41}^{+0.56}$&$0.25_{-0.05}^{+0.07}$&\nodata\\
\hline
11&$1.88_{-1.88}^{+7.82}$&$0.62\pm0.06$&$0.59_{-0.59}^{+0.95}$&$1.27_{-0.75}^{+0.94}$&$0.87_{-0.63}^{+0.93}$&$0.44_{-0.14}^{+0.17}$&\nodata\\
\hline
12a&$<2.30$&$0.60_{-0.09}^{+0.05}$&$0.11_{-0.11}^{+0.48}$&$0.35_{-0.32}^{+0.39}$&$0.20_{-0.20}^{+0.65}$&$0.27_{-0.07}^{+0.09}$&\nodata\\
\hline
12b&$<5.29$&$0.61\pm0.06$&$<0.38$&$0.14_{-0.14}^{+0.65}$&$<0.82$&$0.26_{-0.11}^{+0.12}$&\nodata\\
\hline
13&$0.32_{-0.32}^{+13.04}$&$0.58_{-0.35}^{+0.09}$&$2.57_{-2.57}^{+5.42}$&$1.33_{-1.26}^{+15.18}$&$1.54_{-1.54}^{+136.71}$&$0.47_{-0.30}^{+2.88}$&
$1.88$\tablenotemark{(1)}\\
\hline
14&$3.07_{-3.07}^{+29.18}$&$0.37_{-0.08}^{+0.39}$&$unconstr.$&$unconstr.$&$unconstr.$&$unconstr.$&$1.88$\tablenotemark{(1)}\\
\hline
15&$3.95_{-3.95}^{+10.92}$&$0.59_{-0.14}^{+0.09}$&$1.09_{-1.09}^{+5.52}$&$0.89_{-0.89}^{+2.37}$&$0.37_{-0.37}^{+1.61}$&
$0.72_{-0.34}^{+1.23}$&$1.88$\tablenotemark{(1)}\\
\hline
16&$0.76_{-0.76}^{+4.07}$&$0.62_{-0.05}^{+0.04}$&$0.32_{-0.32}^{+0.46}$&$0.52_{-0.32}^{+0.38}$&$0.56_{-0.34}^{+0.44}$&$0.34_{-0.07}^{+0.08}$&
\nodata\\
\hline
17&$3.08_{-3.08}^{+10.26}$&$0.62_{-0.34}^{+0.10}$&$0.96_{-0.96}^{+14.74}$&$2.91_{-2.24}^{+28.48}$&$1.16_{-1.16}^{+4.94}$&$0.84_{-0.32}^{+3.25}$&
$1.88$\tablenotemark{(1)}\\
\enddata

\tablenotetext{(1)}{Photon index frozen.}

\end{deluxetable}
\clearpage



\begin{deluxetable}{ccccc}
\tablecaption{Best-Fit Parameters (68\% errors) for the Integrated Emission of 
The Antennae (Comparison with {\it ASCA}). 
\label{total}}
\tablewidth{0pt}
\tablehead{
\colhead{$\chi^2/dof$} &
\colhead{\begin{tabular}{c}
$N_H$\\
($\times10^{21}$ cm$^{-2}$)
\end{tabular}} &
\colhead{\begin{tabular}{c}
$kT$\\
(keV)
\end{tabular}} &
\colhead{\begin{tabular}{c}
$Z$\\
($\times Z_{\odot}$)
\end{tabular}} &
\colhead{$\Gamma$}
}
\startdata
1043.5/371 & $0.49_{-0.19}^{+0.12}$ & $0.59\pm0.03$ & $0.18_{-0.08}^{+0.15}$ & 
$1.52_{-0.09}^{+0.17}$\\
\hline
\enddata

\end{deluxetable}
\clearpage


\end{document}